\documentclass[twoside,twocolumn,9pt]{article}
\usepackage{extsizes}
\usepackage[super,sort&compress,comma]{natbib}
\usepackage[version=3]{mhchem}
\usepackage[left=2.0cm, right=2.0cm, top=1.785cm, bottom=2.0cm]{geometry}
\usepackage{balance}
\usepackage{mathptmx}
\usepackage{sectsty}
\usepackage{graphicx}
\usepackage{lastpage}
\usepackage[format=plain,justification=justified,singlelinecheck=false,font={stretch=1.125,small,sf},labelfont=bf,labelsep=space]{caption}
\usepackage{float}
\usepackage{fancyhdr}
\usepackage{fnpos}
\usepackage[english]{babel}
\addto{\captionsenglish}{%
  
}
\usepackage{array}
\usepackage{droidsans}
\usepackage{charter}
\usepackage[T1]{fontenc}
\usepackage[usenames,dvipsnames]{xcolor}
\usepackage{setspace}
\usepackage[compact]{titlesec}
\usepackage{hyperref}
\usepackage{multirow}

\usepackage{dcolumn}

\usepackage{url}

\usepackage{epstopdf}

\graphicspath{ {figures/} }

\definecolor{cream}{RGB}{222,217,201}

\newcommand{\Duo}{{\sc Duo}}

\newcommand{\cm}{cm\textsuperscript{-1}}

\newcommand{\ai}{{\it ab initio}}
\newcommand{\Ai}{{\it Ab initio}}

\newcommand{\X}{$X\,{}^{1}\Sigma^{+}$}
\newcommand{\A}{$A\,{}^{1}\Pi$}
\newcommand{\C}{$C\,{}^{1}\Sigma^{-}$}
\newcommand{\D}{$D\,{}^{1}\Delta$}
\newcommand{\E}{$E\,{}^{1}\Sigma^{+}$}
\newcommand{\as}{$a\,{}^{3}\Sigma^{+}$}
\newcommand{\bp}{$b\,{}^{3}\Pi$}
\newcommand{\dd}{$d\,{}^{3}\Delta$}
\newcommand{\es}{$e\,{}^{3}\Sigma^{-}$}

\newcommand{\allstates}{\X, \A, \C, \D, \E, \as, \bp, \dd, \es}

\newcommand{\alltriplets}{\as, \bp, \dd, \es}

\begin{document}

\title{Rovibronic spectroscopy of PN from first principles$^\dag$}

 \author{Mikhail Semenov$^{a,c}$, Nayla El-Kork$^b$,  Sergei N. Yurchenko$^{a\ast}$ and Jonathan Tennyson$^a$}

\maketitle

\section*{Abstract}

We report an \ai\ study on the rovibronic spectroscopy of the closed-shell diatomic molecule phosphorous mononitride, PN. The study considers the nine lowest electronic states, $X\,{}^{1}\Sigma^{+}$, $A\,{}^{1}\Pi$, $C\,{}^{1}\Sigma^{-}$, $D\,{}^{1}\Delta$, $E\,{}^{1}\Sigma^{-}$, $a\,{}^{3}\Sigma^{+}$, $b\,{}^{3}\Pi$, $d\,{}^{3}\Delta$ and $e\,{}^{3}\Sigma^{-}$ using high level electronic structure theory and accurate nuclear motion calculations. The \textit{ab initio} data cover 9 potential energy, 14 spin-orbit coupling, 7 electronic angular momentum coupling, 9 electric dipole moment and 8 transition dipole moment curves.
The \textsc{Duo} nuclear motion program is used to solve the coupled nuclear motion Schr\"{o}dinger equations for these nine electronic states and to simulate rovibronic absorption spectra of $^{31}$P$^{14}$N for different temperatures, which are compared to  available spectroscopic studies. Lifetimes for all states are calculated and compared to previous results from the literature. The calculated lifetime of the \A\ state shows good agreement with an experimental value from the literature, which is an important quality indicator for the \textit{ab initio} $A$--$X$ transition dipole moment.


\footnotetext{$^a$\textit{Department of Physics and Astronomy, University College London, Gower Street, WC1E 6BT London, United Kingdom}}

\footnotetext{$^{b}$\textit{Department of Physics, Khalifa University, Abu Dhabi, UAE}}

\footnotetext{$^{c}$\textit{Department of Science and Research, Moscow Witte University, 2nd Kozhukhovskiy passage,  Moscow, Russian Federation}}


\footnotetext{$^{\ast}$~Corresponding author; E-mail:s.yurchenko@ucl.ac.uk}
\footnotetext{\dag~Electronic Supplementary Information (ESI) available. See DOI: \href{https://doi.org/10.1039/D1CP02537F}{10.1039/d1cp02537f}}



\date{\today}



\section{Introduction}\label{sec:intro}

Phosphorus is considered to be one of the key elements as a source of life and replication on our planet, \citep{05PaLaxx.PN,20GrRiBa.PH3}  with PN being one of the candidates in star and meteorite evolution to provide the necessary life building material. There have been multiple observations of PN in different media in space: hot dense molecular clouds,\cite{87TuBaxx.PN,87Ziurys.PN} energetic star forming regions,\cite{90TuTsBa.PN,11YaTaSa.PN} cold cloud cores,\cite{90TuTsBa.PN,16FoRiCa.PN} red giant stars \cite{08MiHaTe.PN,13DeKaPa.PN,18ZiScBe.PN} and protoplanetary nebula.\cite{08MiHaTe.PN} Regardless of high astrophysical and astrobilogical importance, phosphorous mononitride is one of the experimentally least-well studied  diatomic molecules of its isoelectronic group (P$_2$, SiO, N$_2$, CS).

The molecule was reported for the first time by \citet{33CuHeHe1.PN} using a spectroscopic study. Many other subsequent spectroscopic studies have taken place  including  photoelectron,\cite{75WuFexx.PN}  fluorescence,\cite{73MoSixx.PN} matrix infrared,\cite{77AtTixx.PN} microwave \cite{72HoTiTo.PN, 72WyGoMa.PN, 06CaClLi.PN} and Fourier Transform Infra-Red (FTIR).\cite{95AhHaxx.PN} Most of the high resolution spectroscopy experiments with PN concentrated on the electronic system   \A\ -- \X,\cite{33CuHeHe1.PN,73MoSixx.PN,81GhVeVa.PN,87SaKrxx.PN,96LeMeDu.PN} with \E\ -- \X\ being confirmed afterwards,\cite{80CoPrxx.PN, 81CoPrxx.PN, 84VeGhxx.PN} and several valence and Rydberg states have been studied as well.\cite{87VeGhIq.PN, 92BrDuMa.PN}


Measured lifetimes can be important indicators of intensities and Einstein A coefficients. So far only one experimental work reports lifetime measurements for the \A\ state of PN molecule \cite{75MoMcSi.PN} using Hanle effect, with several \ai\ works providing computed values for PN lifetimes.\cite{93deFeKo.PN,19QiZhLi.PN}


The mass spectrometric experiment by \citet{69Gingeric.PN} reports PN's dissociation energy $D_0$ to be 6.35$\pm$0.22 eV, which is lower than experimental $D_0$  values by \citet{54HuTaEl.PN} and \citet{68UyKoCa.PN},  7.1$\pm$0.05 eV and 7.57$\pm$0.03 eV. In a combination of a high-level \ai\ and microwave spectroscopy study, \citet{06CaClLi.PN} suggested a $D_0$ value of 6.27~eV. This  is close both to the originally predicted value by \citet{33CuHeHe.PN} and to the  experimental value of \citet{69Gingeric.PN}.

Several theoretical investigations of PN are available in the literature. The most recent was carried out by \citet{19QiZhLi.PN} who reported  spectroscopic constants for the lowest five singlet (\X, \A, \D, \C, $2^{1}\Pi$), six  triplet (including \alltriplets) and two quintet electronic states of PN. In that paper all the states were studied at the internally contracted multi-reference configuration interaction (icMRCI) level of theory with Davidson correction (+Q). Similarly, \citet{14AbSaMa.PN} reported an \ai\ study of seven states of PN, \X , \A , \D , \C , $2^{1}\Pi$, \E, $3^{1}\Sigma^{+}$, as well as  of 10 triplet and 3 quintet electronic states. The main purpose of their study was to interpret perturbation and predissociation effects in the observed transitions.

In this work we  present a comprehensive \ai\ spectroscopic model for the nine lowest electronic states of phosphorous mononitride, \allstates, consisting of  potential energy curves (PECs), transition dipole moments curves (TDMCs), spin-orbit coupling curves (SOCs) and angular momentum coupling curves (AMCs) using the icMRCI+Q method  and to calculate the rovibronic energies and transition probabilities as an \ai\ line list for PN. Producing such line lists for molecules of astrophysical significance is one of the main objectives of the ExoMol project.\cite{jt528}  These curves, with some simple adjustment of the minimum energies of the PECs, are used to solve the coupled nuclear-motion Schr\"{o}dinger equation with the program \Duo.\cite{jt609} The spectroscopic model and \ai\ curves are provided as part of the supplementary material. Our open source code \Duo\ can be accessed via \url{http://exomol.com/software/}. The results of \Duo\ calculation are then used to generate rovibronic spectra of PN and compare to systems previously reported in the literature.

\section{Computational details}

\subsection{\Ai\ calculations}

Using MOLPRO 2020, \cite{MOLPRO2020}  \ai\ calculations were performed for nine low-lying electronic states of PN. Apart from PECs, we also computed  (T)DMCs, SOCs  and EAMCs. For the \ai\ calculations  we used  the icMRCI method  \cite{89Dunning.ai}  in conjunction with the effective-core-potential (ECP) method ECP10MWB (Stuttgart/Cologne)  for phosphorus~\citep{05LiScMe.ai} and aug-cc-pCV5Z~\citep{93WoDuxx.ai} basis set for nitrogen.
The initial complete active space self-consistent field (CASSCF) calculation over which the configuration interaction calculations were built was for the \X\ state only.  In conjunction with ECP for phosphorus, the active space was selected to be (6,2,2,0) with  (1,0,0,0) closed orbitals.
The state averaging set contained 96 states: 24, 24, 24 and 24 states of the singlets, triplets, pentets and septets. This level of theory will be referenced to as icMRCI+Q/ECP10MWB.

The calculations of the spin-orbit couplings  were too difficult to perform for the icMRCI+Q/ECP10MWB level of theory, taking too long to complete and producing wrong data.
We therefore decided to use a non-ECP level of theory for our spin-orbit calculations. To this end, we selected the \ai\  level of theory similar to that  used by \citet{19QiZhLi.PN}  with an  active space of (9,3,3,0). In this case the state averaging set consisted of 11 singlet configurations (4 $A_1$, 2 $B_1$, 2 $B_2$ and 3 $A_2$). The Douglass-Kroll correction was taken into account with or without core-valence correlation. These levels of theory will be referenced to as icMRCI/aug-cc-pV5Z(-DK) and icMRCI/aug-cc-pWCV5Z(-DK), respectively, with or without DK.


Figures \ref{f:PECsall}--\ref{f:dmcs_abinito} show all PECs, SOCs, EAMSc and TDMCs generated in this study.
If the MOLPRO calculations at some geometries did not converge, they were interpolated or extrapolated from the surrounding points as part of the Duo calculations (see below). We used an adaptive  \ai\ grid  consisting of 150 bond lengths ranging from 0.7 to 8~\AA\  with more points around the equilibrium region. The grid points with the corresponding  \ai\ values (if converged) are included in the supplementary material and also shown in Figs.~\ref{f:PECsall}--\ref{f:dmcs_abinito}. The icMRCI/aug-cc-pV5Z(-DK) calculations have only converged up to about 3.5~\AA\ due to a smaller number of configurations in the internally contacted reference set used.

\begin{table*}
    \caption{Comparison of spectroscopic constants taken from previous works and calculated from our \ai\ curves (icMRCI+Q/ECP10MWB): Dissociation energy $D_{\rm e}$ in \cm (rounded to 3 s.f.), electronic equilibrium energy  $T_{\rm e}$ in \cm, equilibrium bond length $r_{\rm e}$ in \AA, harmonic constant $\omega$  in \cm, rotational constant $B_{\rm e}$ in \cm. }
    \label{t:spectroscopic_constants}
    \begin{tabular}{ l l r rllll}
    \hline
        State &  & $D_{\rm e}$ & $T_{\rm e}$ & $r_{\rm e}$ & $\omega_{\rm e}$ &  $B_{\rm e}$ &  \\ \hline
        \multirow{6}{0.1em}{\X} & This work & 53100.00 & 0.00 & 1.49 & 1303.23 & 0.78 &  \\
         & Expt.\cite{33CuHeHe.PN} & 50812.91 & 0.00 & 1.4869 & 1337.24 & 0.78549 &  \\
         & Expt.\cite{06CaClLi.PN} & 51938.86 & 0.00 &  & 1336.992 & 0.78648 &  \\
         & Calc \cite{19QiZhLi.PN} & 51454.07 & 0.00 & 1.4918 & 1339.61 &  0.78549 &  \\
         & Calc. \cite{10WaJiDe.PN} & 50645.15 & 0.00 & 1.4948 & 1333.84 &  0.7823 &  \\
         & Calc. \cite{14AbSaMa.PN} & 49683.73 & 0.00 & 1.4977 & 1328.21 &  0.77872 &  \\ \hline
        \multirow{4}{0.1em}{\A} & This work & 44900.00 & 42143.66 & 1.56 & 1041.07 &  0.71 &  \\
         & Expt. \cite{33CuHeHe.PN} & 41134.26 & 39688.52 & 1.5424 & 1103.09 &  0.73071 &  \\
         & Expt. \cite{81GhVeVa.PN} &  & 39805.90 &  & 1103 &  0.731 &  \\
         & Calc \cite{19QiZhLi.PN} & 40995.53 & 40032.55 & 1.5476 & 1104.81 & 0.7306 &  \\
         & Calc. \cite{10WaJiDe.PN} & 41331.86 &  & 1.5501 & 1100.24 & 0.727505 &  \\
         & Calc. \cite{14AbSaMa.PN} & 40624.26 & 40610.00 & 1.556 & 1078.42 & 0.7213 &  \\ \hline
        \multirow{5}{0.1em}{\D} & This work & 43100.07 & 43287.03 & 1.63 & 1016.44  & 0.66 &  \\
         & Calc \cite{19QiZhLi.PN} & 39993.49 & 42106.86 & 1.6073 & 1018.98 &  0.67724 &  \\
         & Calc. \cite{83GrKaxx.PN} & 33713.96 & 41618.19 & 1.622 & 967.9 &  0.664 &  \\
         & Calc. \cite{14AbSaMa.PN} & 39448.89 & 41835.96 & 1.614 & 1010.647 & 0.671 &  \\
         & Calc. \cite{10WaJiDe.PN} & 41555.28 &  & 1.6196 & 1002.269 &  0.665056 &  \\ \hline
        \multirow{4}{0.1em}{\C} & This work & 45300.00 & 41417.18 & 1.63 & 998.65 &  0.66 &  \\
         & Calc \cite{19QiZhLi.PN} & 42766.02 & 39298.72 & 1.6211 & 977.5 & 0.6639 &  \\
         & Calc. \cite{83GrKaxx.PN} & 37827.38 & 37504.77 & 1.617 & 1108.8 &  0.669 &  \\
         & Calc. \cite{14AbSaMa.PN} & 41852.09 & 39505.02 & 1.627 & 973.927 &  0.6597 &  \\ \hline
        \multirow{2}{0.1em}{\E} & This work & 30200.00  & 56793.82 & 1.86 & 702.48 &  0.50 &  \\
         & Calc. \cite{14AbSaMa.PN} & 26188.80 & 55103.80 & 1.889 & 743.74 &  0.4893 &  \\ \hline
        \multirow{4}{0.1em}{\as} & This work & 26600.62 & 26632.70 & 1.66 & 917.51 &  0.63 &  \\
         & Calc \cite{19QiZhLi.PN} & 25249.32 & 26065.66 & 1.6481 & 941.61 &  0.64253 &  \\
         & Calc. \cite{83GrKaxx.PN} & 19599.26 & 24761.21 & 1.669 & 787.4 &  0.628 &  \\
         & Calc. \cite{14AbSaMa.PN} & 23785.28 & 25817.80 & 1.655 & 913.765 &  0.6375 &  \\ \hline
        \multirow{3}{0.1em}{\es} & This work & 32670.00 & 39370.41 & 1.64 & 974.15 &  0.65 &  \\
         & Calc \cite{19QiZhLi.PN} & 32081.01 & 37951.02 & 1.624 & 981.37 &  0.66149 &  \\
         & Calc. \cite{83GrKaxx.PN} & 37827.39 & 37504.77 & 1.617 & 1108.8 &  0.669 &  \\
         & Calc. \cite{14AbSaMa.PN} & 30665.19 & 37956.44 & 1.632 & 982.255 &  0.6556 &  \\ \hline
        \multirow{4}{0.1em}{\dd} & This work & 33300.00 & 34860.41 & 1.65 & 960.01 &  0.65 &  \\
         & Calc \cite{19QiZhLi.PN} & 29679.37 & 32975.56 & 1.6324 & 949.99 &  0.65376 &  \\
         & Calc. \cite{83GrKaxx.PN} & 23551.38 & 32504.13 & 1.666 & 770.3 & 0.629 &  \\
         & Calc. \cite{14AbSaMa.PN} & 28471.36 & 33125.18 & 1.64 & 959.263 &  0.6495 &  \\ \hline
        \multirow{4}{0.1em}{\bp} & This work & 33400.00 & 35146.16 & 1.56 & 1070.81 &  0.71 &  \\
         & Calc \cite{19QiZhLi.PN} & 29197.77 & 33671.79 & 1.5449 & 1113.24 & 0.73301 &  \\
         & Calc. \cite{83GrKaxx.PN} & 21696.31 & 34359.20 & 1.558 & 1124.3 &  0.72 &  \\
         & Calc. \cite{14AbSaMa.PN} & 27672.87 & 33843.01 & 1.555 & 1102.494 &  0.7224 &  \\ \hline
    \end{tabular}

\end{table*}

\subsection{\Duo\ calculations}

We use the program \Duo\ \cite{jt609,jt626} to solve the coupled Schr\"{o}dinger equation for 9 lowest electronic states of PN. \Duo\ is a variational program capable of solving rovibronic problems for a general (open-shell) diatomic molecule with an arbitrary number of couplings, see, for example,
Refs.~\citenum{jt589,jt736,jt759,jt760}. All \ai\ couplings between these 9 states are taken into account as described below.  The goal of this paper is to build a comprehensive \ai\ spectroscopic model for this  electronic  system of PN based on the icMRCI+Q/ECP10MWB and icMRCI/aug-cc-pV5Z-DK \ai\ curves. We therefore do not attempt a systematic refinement of the \ai\ curves by fitting to the experiment, which will be the subject of future work. In order to facilitate the comparison with the experimental data, we, however, perform some shifts of the equilibrium energy $T_{\rm e}$ and bond length $r_{\rm e}$ values, as described in further detail below. We also smooth some of the \ai\ curves as described in detail in the Appendix.

In \Duo\ calculations, the coupled Schr\"{o}dinger equations are solved on an equidistant grid of points, in our case 501, with bond lengths $r_i$ ranging from  $r = $ 0.85 to 5 \AA\ using the sinc DVR method \cite{82GuRoxx}. Our \ai\ curves are represented by sparser and less extended grids (see above). For the bond length values $r_i$ overlapping with the \ai\ ranges, the \ai\ curves were projected onto the denser \Duo\ grid using the cubic spline interpolation.

The following functional forms were used for  extrapolation outside the original \ai\ range:\citep{jt609}
\begin{eqnarray}
\nonumber
  f_{\rm PEC}^{\rm short}(r) &=& A + B/r, \\
  f_{\rm TDMC}^{\rm short}(r) &=& A r + B r^2,\\
  \nonumber
  f_{\rm other}^{\rm short}(r) &=& A + B r,
\end{eqnarray}
for short range and
\begin{eqnarray}
\nonumber
 f_{\rm PEC}^{\rm long}(r)  &=& A + B/r^6\\
 f_{\rm EAMC}^{\rm long}(r)  &=& A + B r, \\
 \nonumber
 f_{\rm other}^{\rm long}(r)  &=& A/r^2 + B/r^3
\end{eqnarray}
for  long range, where $A$ and $B$ are stitching parameters.

The vibrational basis set was taken as eigensolutions of the nine uncoupled 1D problems for each PEC.  The corresponding basis set constructed from 9$\times $501 eigenfunctions was then contracted to include about 60--80 vibrational functions from each  state, enough to fill the PEC up to the corresponding  dissociation energies. This gives a total of 640 states. These vibrational basis functions were then combined with the spherical harmonics for the rotational and electronic spin basis set functions. All calculations were performed for $^{31}$P$^{14}$N using atomic masses.

\section{Results and discussion}

\subsection{Results of \ai\ calculations}

\begin{figure}
    \includegraphics[width=0.9\columnwidth]{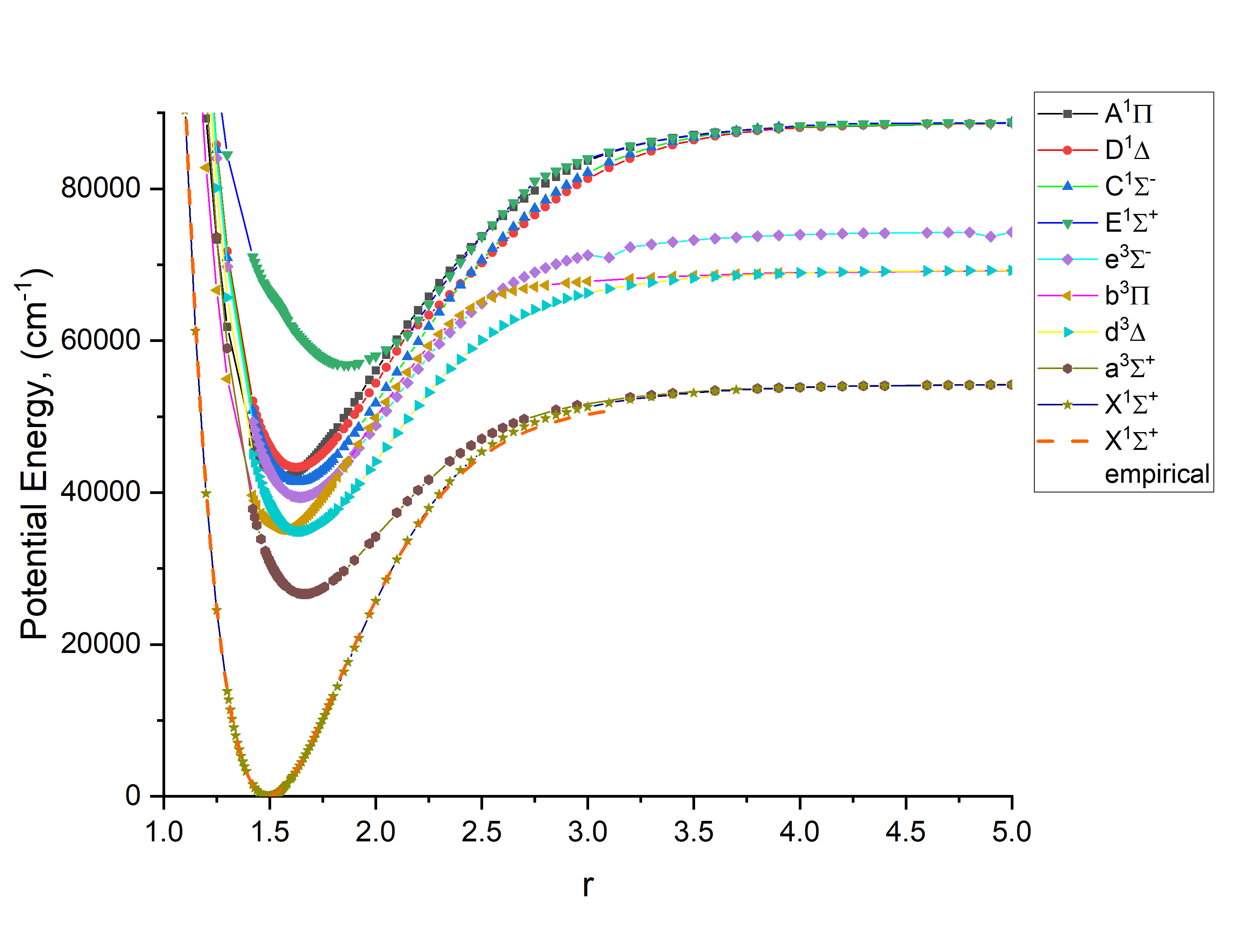}
	\caption{icMRCI+Q/ECP10MWB level calculated PECs of PN: The lowest 5 singlets and 4 lowest triplets. }
	\label{f:PECsall}
\end{figure}

The lowest 9 singlet and triplet  PECs (\allstates)  calculated at the icMRCI+Q/ECP10MWB level and are shown in Fig.~\ref{f:PECsall}.  Table~\ref{t:spectroscopic_constants} presents spectroscopic constants for these states estimated using the corresponding (spin-orbit-free) PECs  and compares to  previous studies \cite{33CuHeHe.PN,06CaClLi.PN, 19QiZhLi.PN,10WaJiDe.PN, 14AbSaMa.PN, 81GhVeVa.PN, 83GrKaxx.PN}. These values agree well for the  states \X, \C\, \D, \E, \as, \es, \bp, \dd\ as estimated using the corresponding (spin-orbit-free) PECs showing a reasonably good agreement. For the \E\ state, however we note that there are very few studies of that state, so further computational investigation is potentially required. For the \A\ state, the  $T_{\rm e}$ value (the equilibrium electronic energy relative to the minimum value of \X) is by  about 2300 \cm\ larger than previous experimental  and \ai\ values, which in turn affects the $D_{\rm e}$ value. We consider this to be an effect of the ECP method  used in the \ai\ calculations and we note that there is a similar magnitude shift in $T_{\rm e}$  of all other states.

\begin{figure*}
  \includegraphics[width=0.95\columnwidth]{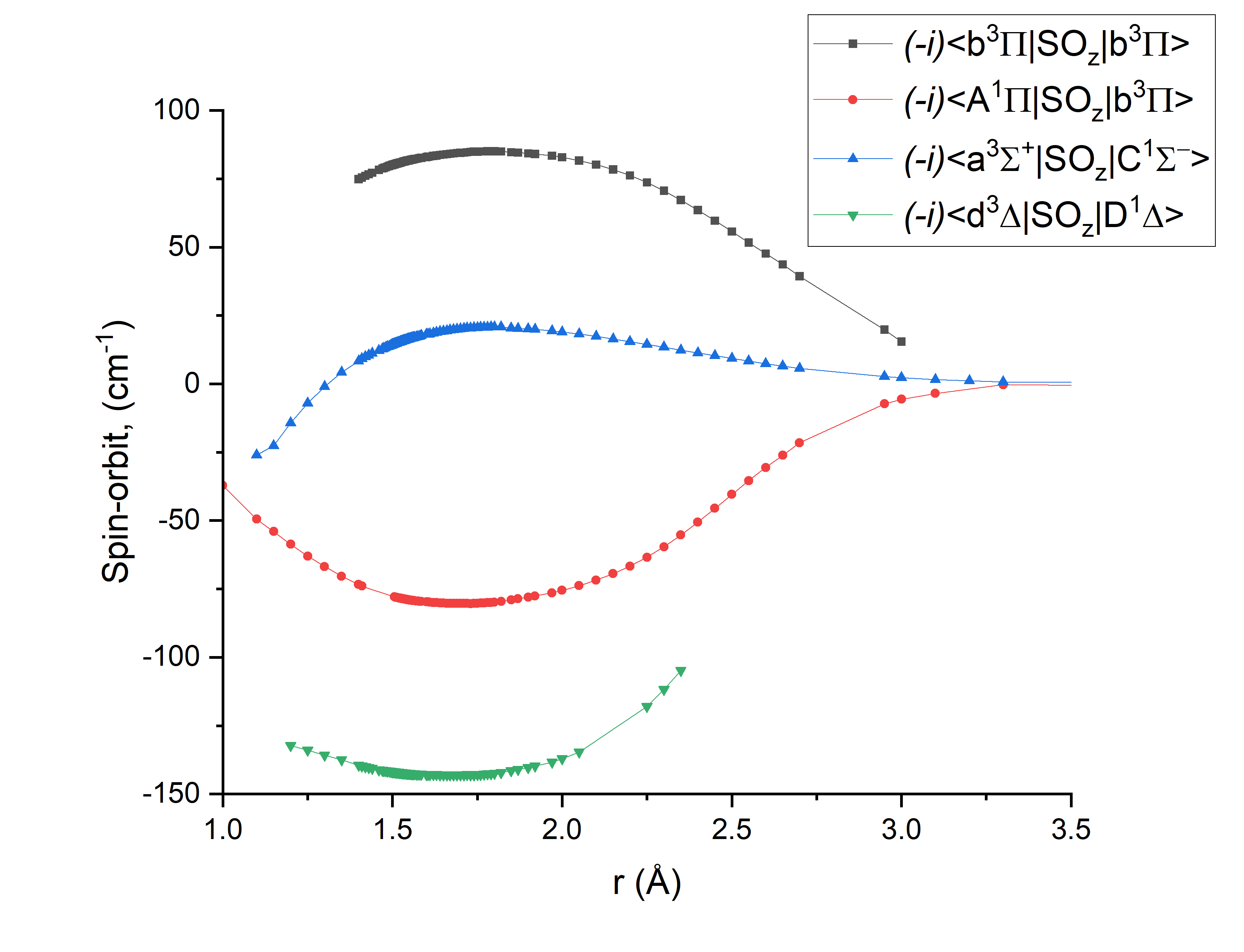}
    \includegraphics[width=0.95\columnwidth]{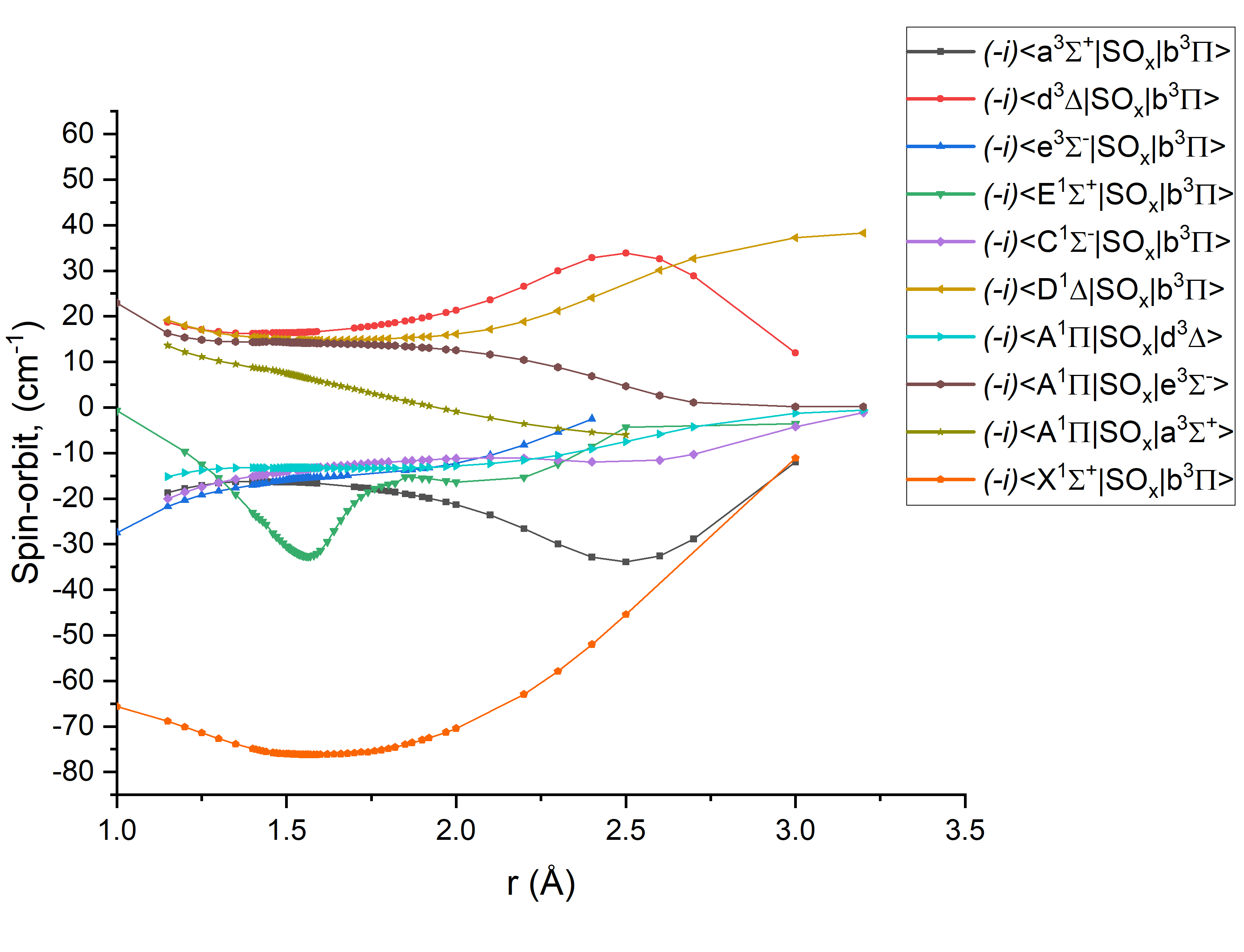}
	\caption{Calculated   spin-orbit matrix elements $\langle i$|SO$_x|j \rangle$ for PN at the 
	icMRCI/aug-cc-pV5Z-DK level of theory. The MOLPRO values of the magnetic quantum numbers $m_S$ for the curves can be found in Table.~\ref{t:mS_values}.}
  \label{f:socs}
\end{figure*}

\begin{figure}
	\includegraphics[width=0.8\columnwidth]{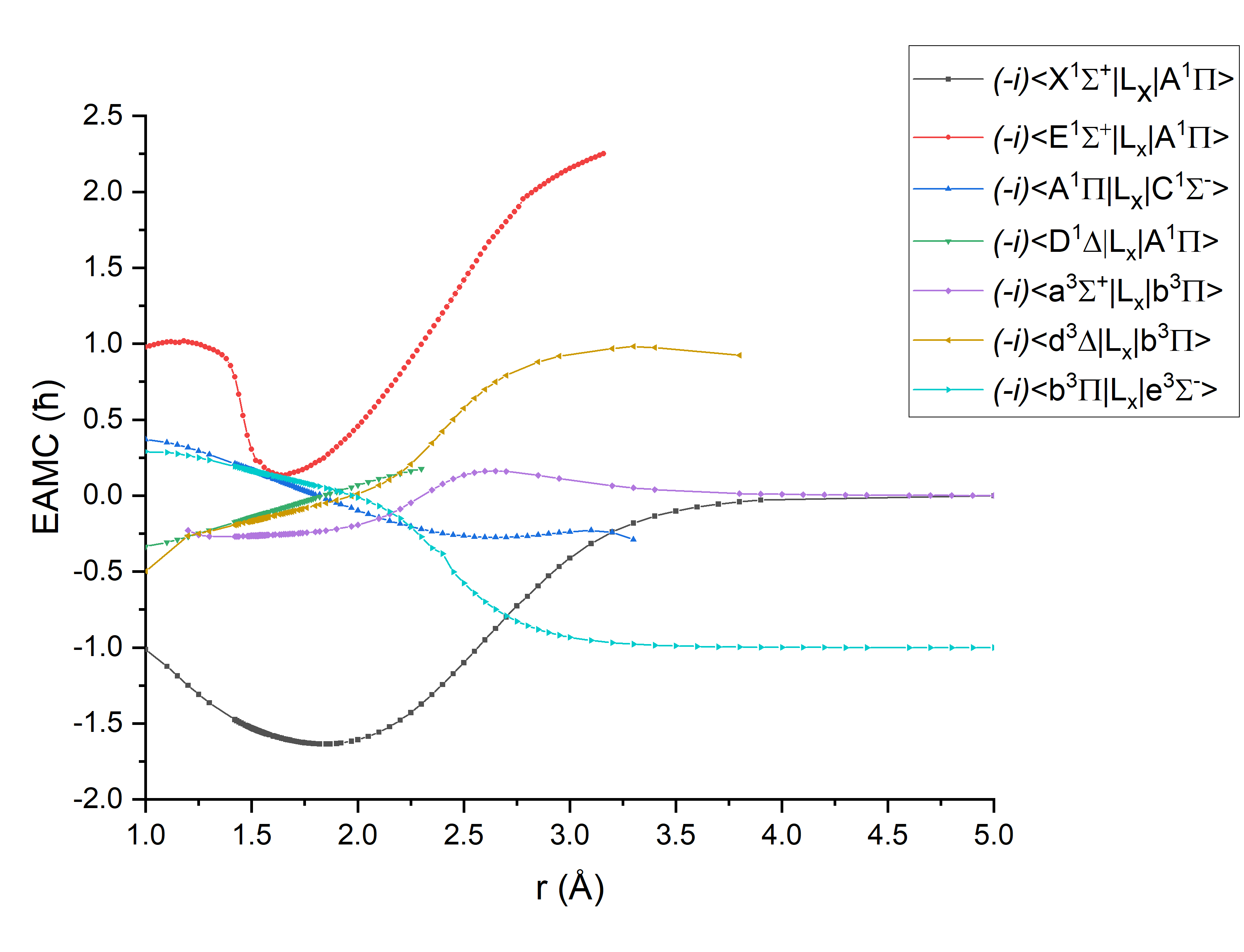}
	\caption{\textit{Ab initio} (icMRCI+Q/ECP10MWB)  electronic angular momentum couplings for PN in the units of $\hbar$.}
	\label{f:ang_mom_abinitio}
\end{figure}

\begin{figure}
    \centering
    \begin{minipage}{0.9\columnwidth}
    \centering
	\includegraphics[width=0.99\columnwidth]{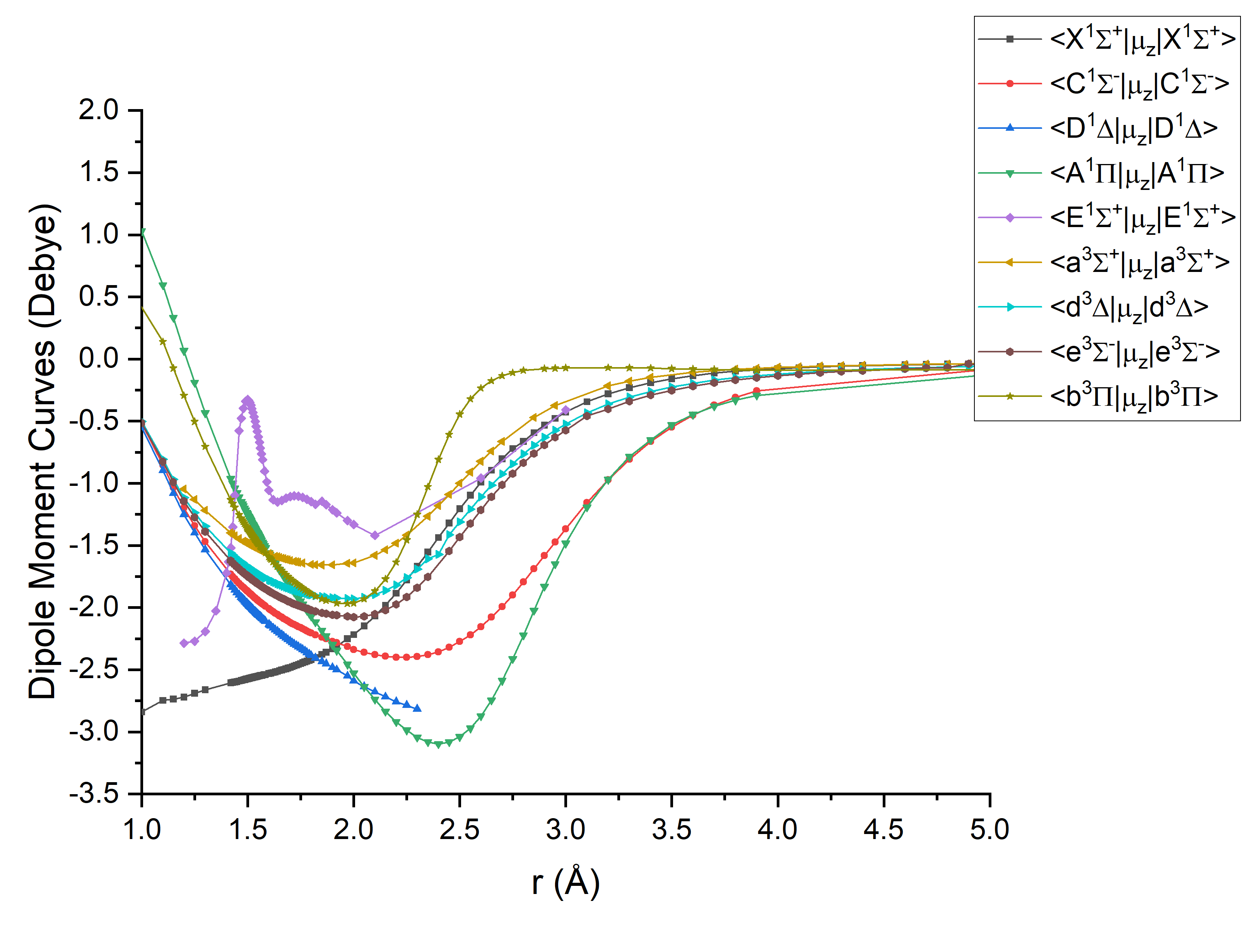}
	\end{minipage}
	\begin{minipage}{0.8\columnwidth}
    \centering
	\includegraphics[width=0.99\columnwidth]{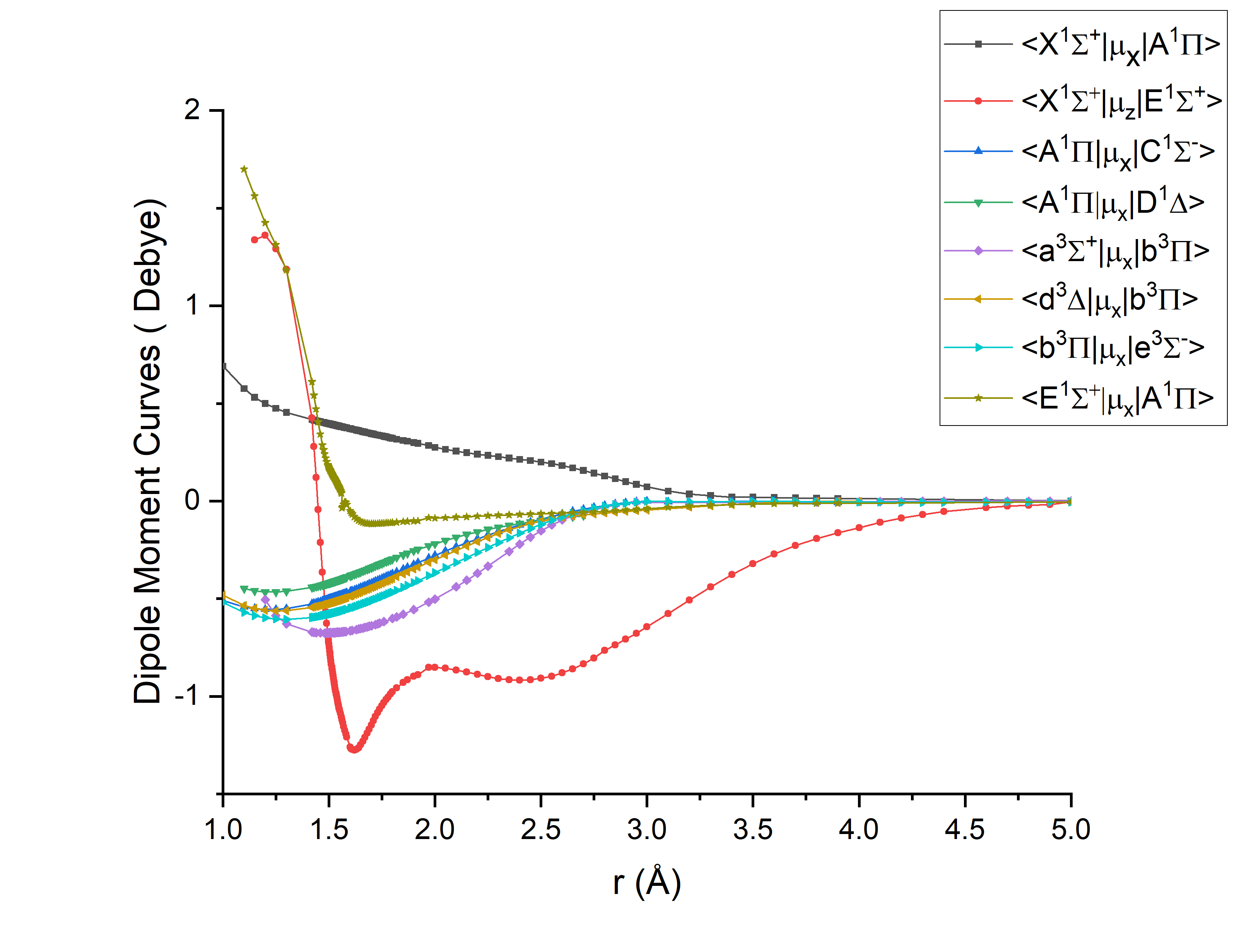}
	\end{minipage}
	\caption{\textit{Ab initio} (icMRCI+Q/ECP10MWB) dipole moment curves for PN: diagonal (upper), off-diagonal (lower).}
	\label{f:dmcs_abinito}
\end{figure}

The \ai\ SOCs, EAMCs and TDMCs are shown in Figures ~\ref{f:socs}, ~\ref{f:ang_mom_abinitio}, ~\ref{f:dmcs_abinito} respectively, with their phases indicated and the reference distance for phases taken as 1.5 \AA.  For all these curves the post-processing of \ai\ calculations included (i) inter- and extrapolations for the missing points using neighboring geometries and (ii) phase mapping using a procedure similar to the one described by \citet{jt589}. Some of the curves also cover only part of the range, which can be seen with the TDMCs   $\langle$\X$|\mu_z|$\E$\rangle$ in Fig.~\ref{f:dmcs_abinito}. The rest of the points portrayed incoherent behaviour and hence were dropped in the post-processing. When calculating spectra for curves with dropped points, the lines were extrapolated using cubic splines as part of the Duo calculations.

\begin{table}
    \caption{MOLPRO magnetic quantum numbers  $m_S$ values for the \ai\ spin-orbit matrix elements  displayed in Fig.~\ref{f:socs}.  }
    \label{t:mS_values}
    \centering
    \begin{tabular}{r@{}c@{}lcc}
    \hline
    \multicolumn{3}{c}{SOC}& bra $m_S$ & ket $m_S$ \\
    \hline
    $\langle$ \A$_y$|& SO$_x$&|\dd$_{xy}\rangle$ & 0 & 1  \\
    $\langle$ \A$_y$|& SO$_x$&|\es$\rangle$ & 0 & 1  \\
    $\langle$ \as|& SO$_x$&|\bp$_y\rangle$ & 0 & 1  \\
    $\langle$ \as|& SO$_x$&|\A$_y\rangle$ & 1 & 0  \\
    $\langle$ \dd$_{xy}$|& SO$_x$&|\bp$_y\rangle$ & 0 & 1  \\
    $\langle$ \es|& SO$_x$&|\bp$_x\rangle$ & 0 & 1  \\
    $\langle$ \X|& SO$_x$&|\bp$_y\rangle$ & 0 & 1  \\
    $\langle$ \E|& SO$_x$&|\bp$_y\rangle$ & 0 & 1  \\
    $\langle$ \C|& SO$_x$&|\bp$_y\rangle$ & 0 & 1  \\
    $\langle$ \D$_z$|& SO$_x$&|\bp$_x\rangle$ & 0 & 1  \\
    $\langle$ \bp$_x$|& SO$_z$&|\bp$_x\rangle$ &  1 & 1   \\
    $\langle$ \A$_x$|& SO$_z$&|\bp$_x\rangle$ &  0 & 0   \\
    $\langle$ \as|& SO$_z$&|\C$\rangle$ &  0 & 0   \\
    $\langle$ \dd$_z$|& SO$_z$&|\D$_z\rangle$ &  0 & 0   \\
    \hline
    \end{tabular}
 \end{table}

\subsection{ The \A--\X\ band using different levels of \ai\ theory }

\begin{figure}
  \includegraphics[width=0.9\columnwidth]{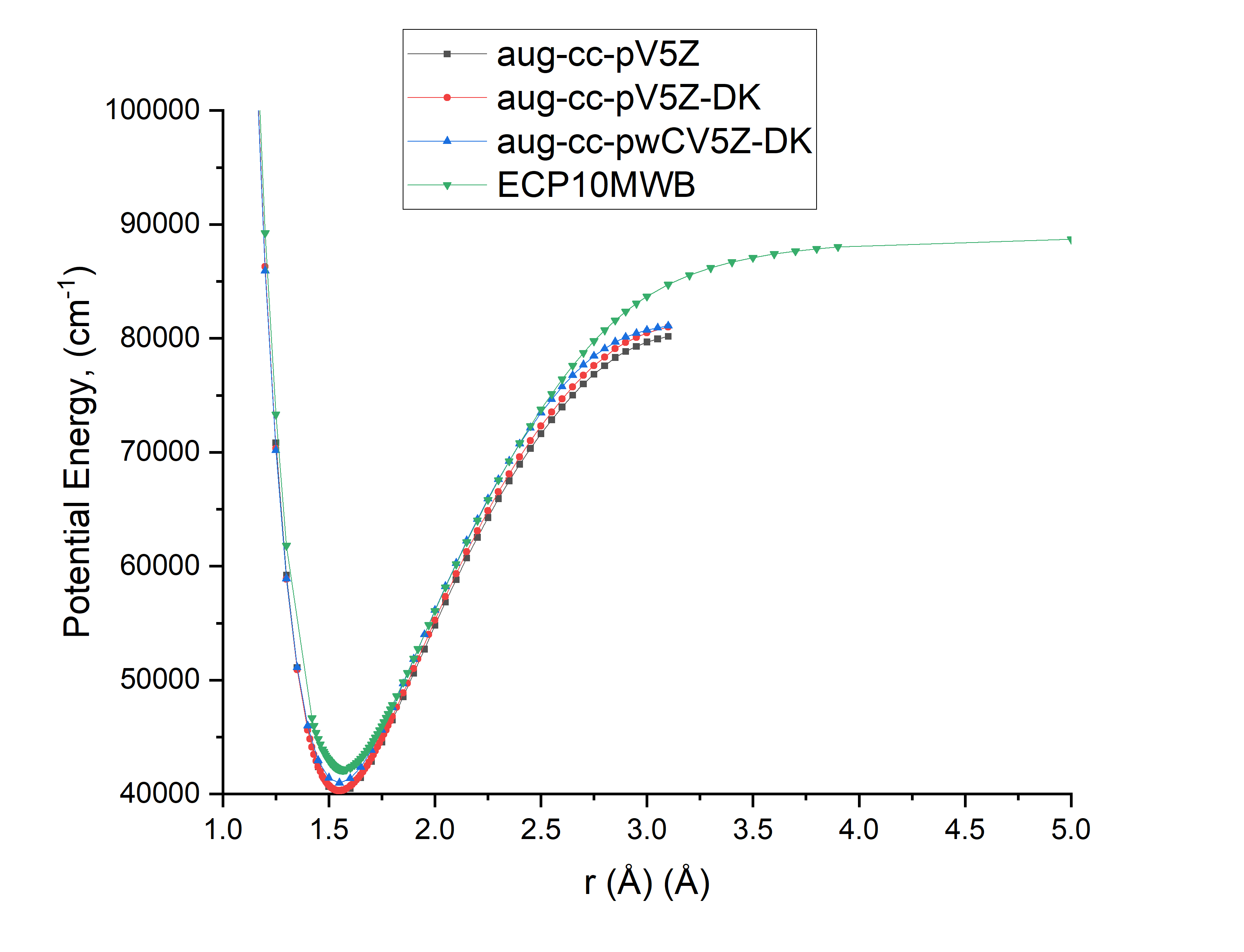}
	\caption{Comparison of PECs for the \A\ state of PN at different levels of theory. }
  \label{f:A1Pi_PEC_Comparison}
\end{figure}

\begin{figure}
  \includegraphics[width=0.9\columnwidth]{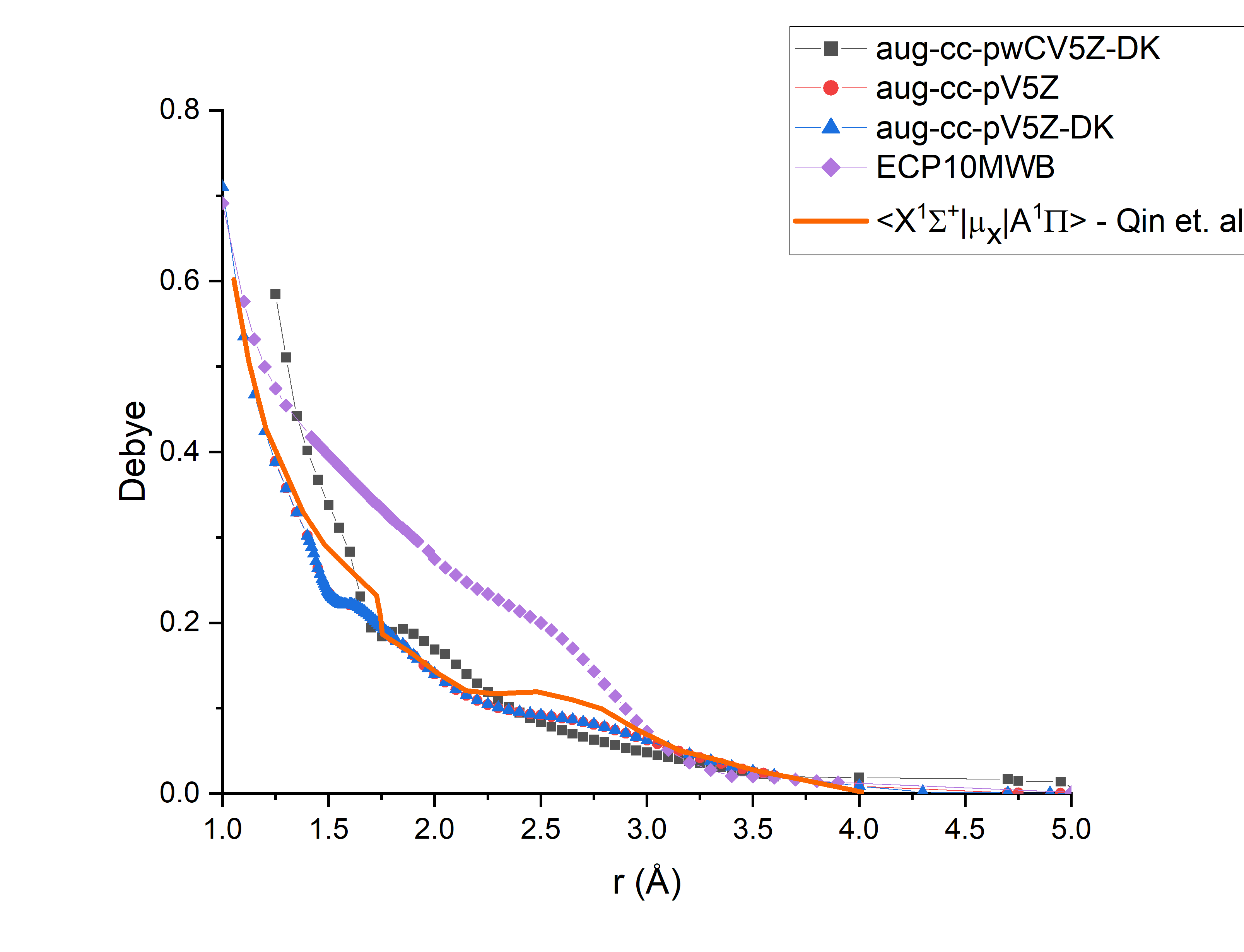}
	\caption{Comparison of transition dipole moment between \A\ and \X\ states of PN at the different levels of theory (ECP10MWB, icMRCI/aug-cc-pV5Z-DK, icMRCI/aug-cc-pWCV5Z-DK) and previously reported results \citep{19QiZhLi.PN}.}
  \label{f:A1Pi_TDMC_comparison}
\end{figure}

As the \A--\X\ band is one of the most important spectroscopic systems of PN, here we compare  the \X\ and \A\ PECs and the \A\ - \X\ TDMC  computed using different levels of theory: icMRCI+Q/ECP10MWB,  icMRCI/aug-cc-pV5Z-DK and icMRCI/aug-cc-pwCV5Z-DK, see Figs.~\ref{f:A1Pi_PEC_Comparison} and \ref{f:A1Pi_TDMC_comparison}.   As part of the spectroscopic model for PN, even a slight change of these curves can have a profound effect on the simulated spectra or lifetimes.


In Figure~\ref{f:A1Pi_TDMC_comparison}, we present a comparison of TDMCs of the \A\ and \X\ states calculated at different levels of theory and results previously calculated by \citet{19QiZhLi.PN}. The TDMC calculated with the ECP10MWB stands out the most with larger values of the dipole moment around the equilibrium (i.e. in the spectroscopically relevant region) and  the rest of the ECP-free methods giving curves which lie close to each other. We show below that the ECP10MWB DMC has  improved  the $A$--$X$ lifetime of PN, the only known experimental evidence of the transition probability of this system.

The aug-cc-pWCV5Z-DK and previous results by \citet{19QiZhLi.PN} show smaller values of TDMC which will lead to weaker intensities and longer lifetimes.


%

\subsection{Results of \Duo\ calculations}

In this study we work directly with the \ai\ data in the grid representation without representing the
curves analytically, with exception of the \X\ state for which we use an empirical potential energy function from \citet{jt590}. Using their PEC, we essentially reproduce the \X\ states ro-vibrational energies of the YYLT line list.

\begin{figure}
    \centering
	\includegraphics[width=0.8\columnwidth]{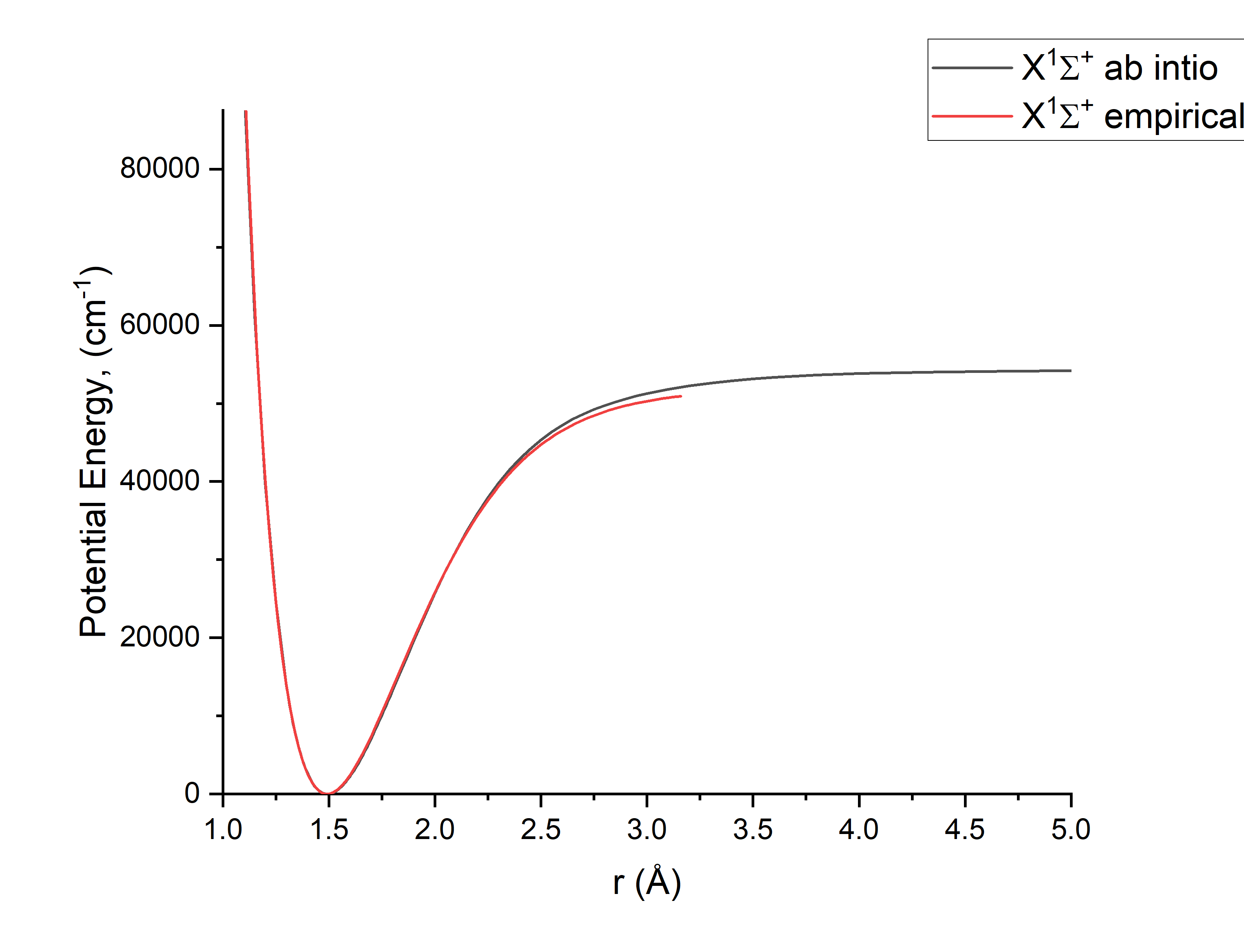}
	\caption{Comparison between a pseudopotential \ai\ \X\ PEC and previously calculated empirical  PEC \cite{jt590}  of PN. }
	\label{f:XPEC_comparison}
\end{figure}

\begin{figure}
    \centering
	\includegraphics[width=0.8\columnwidth]{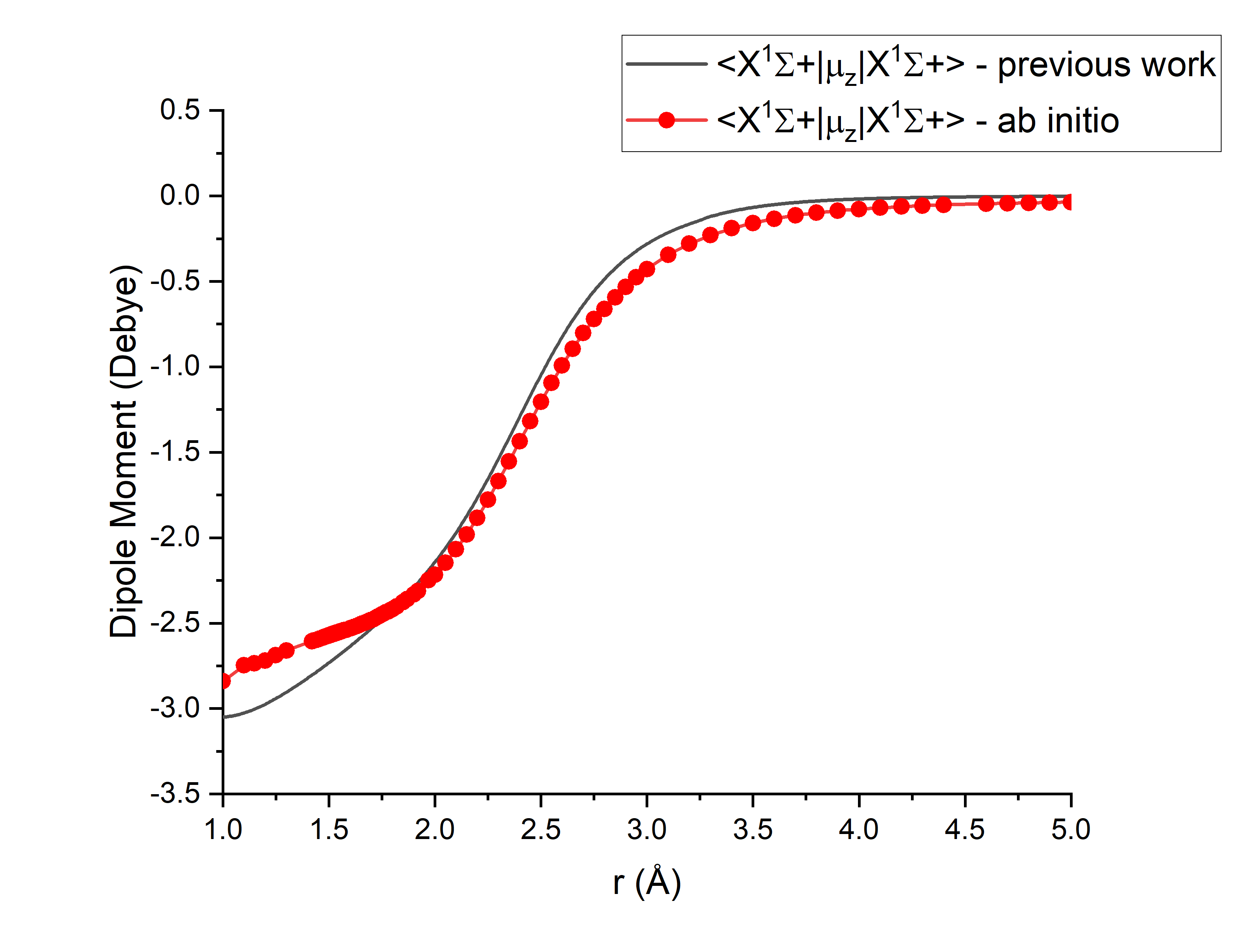}
	\caption{Comparison between absolute values of the \ai\ \X\  DMCs from this work (icMRCI+Q/ECP10MWB) and by  \cite{jt590} (MRCI+Q/aug-cc-pCV6Z)}
	\label{f:Xdipole_comparison}
\end{figure}

For \Duo\ calculations, we selected the following set of curves: the
icMRCI+Q/ECP10MWB PECs shown in Fig.~\ref{f:PECsall} for all but the \X\ state. The comparison between the \X\ \ai\ PEC from this work and the empirical PEC from \cite{jt590} can be seen in Fig.~\ref{f:XPEC_comparison}. Similarly for (T)DMCs all curves are  \ai, see Fig.~\ref{f:dmcs_abinito}, apart from the ground state dipole moment $\langle$\X$|\mu_{z}|$\X$\rangle$, which was  also taken from \citet{jt590}. The comparison between two DMCs is shown in Fig.~\ref{f:Xdipole_comparison}. For SOCs, TDMCs and EAMCs we use \ai\ obtained curves shown in Figs.~\ref{f:socs}, \ref{f:ang_mom_abinitio} and \ref{f:dmcs_abinito}, respectively.

The  rovibronic wave functions of PN were computed with \Duo\ and then used in conjunction with the \ai\ TDMCs to produce Einstein~$A$ coefficients for all rovibronic transitions between states considered in this work, for the wavenumber range from 0 to 88~000~\cm\ and $J \le 270$. In these calculations, the lower states were capped by the energy threshold 60~0000~\cm, which is close to the lowest dissociation threshold, while the upper state threshold was set to 90~000~\cm\ (highest asymptotic channel  of our nine electronic state system). The line list contains 233~418~251 transitions between 390~368 rovibronic states.

These Einstein~$A$ coefficients, organised as per the ExoMol format \citep{jt631} in a line list, were then used to calculate lifetimes and spectra. The aforementioned format uses a two file system to represent relevant spectroscopic information, with the energies and specific state quantum numbers included into the States file (.states) and Einstein coefficients appearing in the Transitions file linking different states (.trans). The .states and .trans files produced by \Duo\ are used in conjunction with \textsc{Exocross} \cite{jt708} to produce spectra and lifetimes (see below).

\subsection{Partition function}

\begin{figure}
    \centering
    \includegraphics[width=0.8\columnwidth]{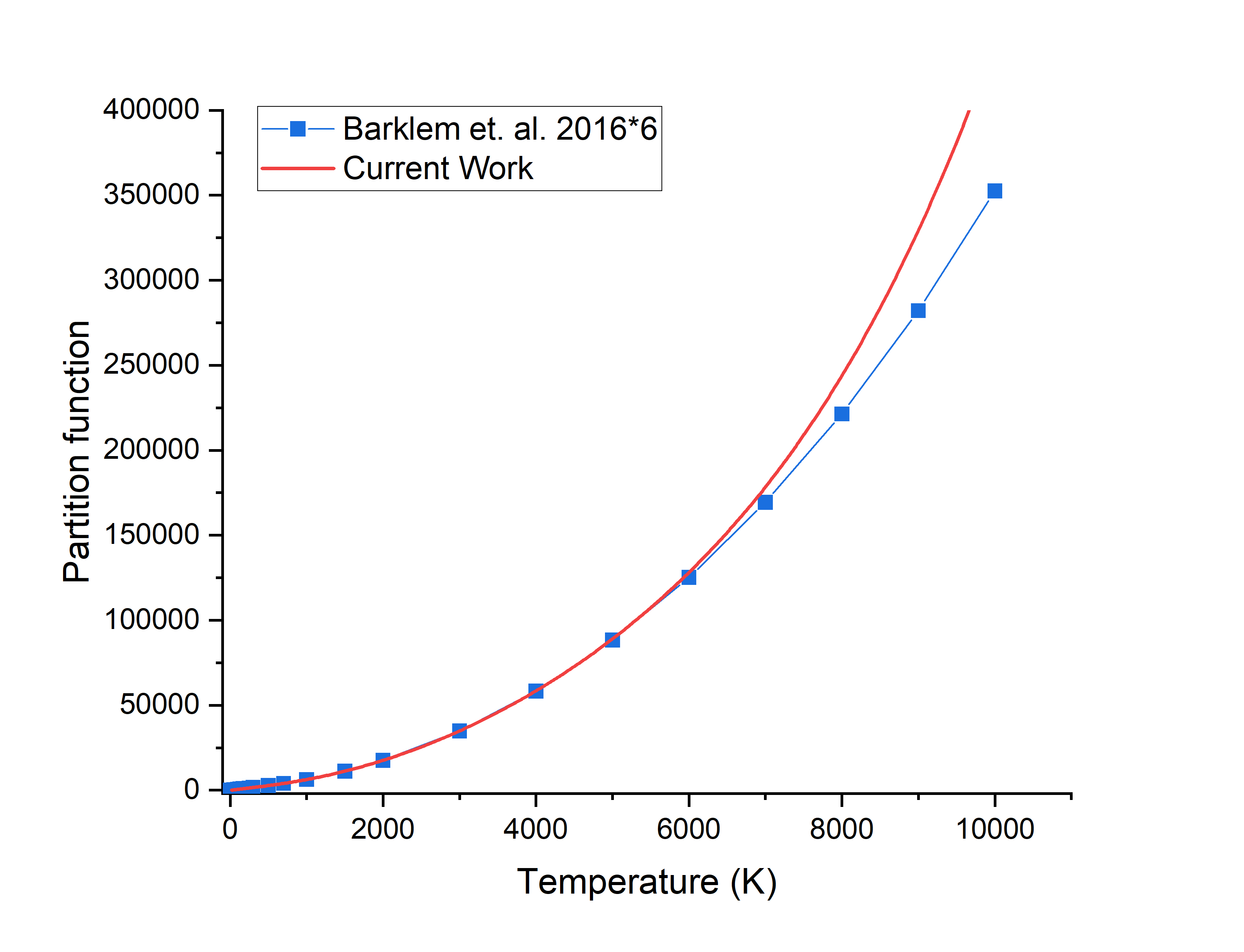}
    \caption{Comparison of partition functions for $^{31}$P$^{14}$N:   \ai\ this work and the values of \citet{16BaCoxx.partfunc}.}
    \label{f:partition_fucntion}
\end{figure}

The partition function of $^{31}$P$^{14}$N computed using our \ai\ line list is shown in Fig.~\ref{f:partition_fucntion},
which is compared to that recently reported by \citet{16BaCoxx.partfunc}. Since $^{14}$N has a nuclear spin
degeneracy of 3 and $^{31}$P has nuclear spin degeneracy of 2, we have multiplied Barklem and Collet's partition function by a factor of six  to compensate for the different conventions used; we follow ExoMol and HITRAN \cite{jt692} and include the full nuclear spin in our partition functions. The differences in higher temperatures can be attributed to incompleteness in the model used by \citet{16BaCoxx.partfunc}.

\subsection{Spectral comparisons}

Using the \ai\ $^{31}$P$^{14}$N line list, spectral simulations were performed with our code
\textsc{ExoCross} \cite{jt708}. \textsc{ExoCross} is an open source Fortran 2003 code,  see \verb!http://exomol.com/software/! or \verb!https://github.com/exomol!,  whose  primary use is to produce spectra of molecules at different temperatures and pressures in the form of cross sections using molecular line lists as input.   Amongst other features, \textsc{ExoCross} can generate spectra for non-local thermal equilibrium conditions characterized with different vibrational and rotational temperatures, lifetimes, Land\'e $g$-factors,
partition and cooling functions.

\begin{figure}
    \centering
    \includegraphics[width=0.9\columnwidth]{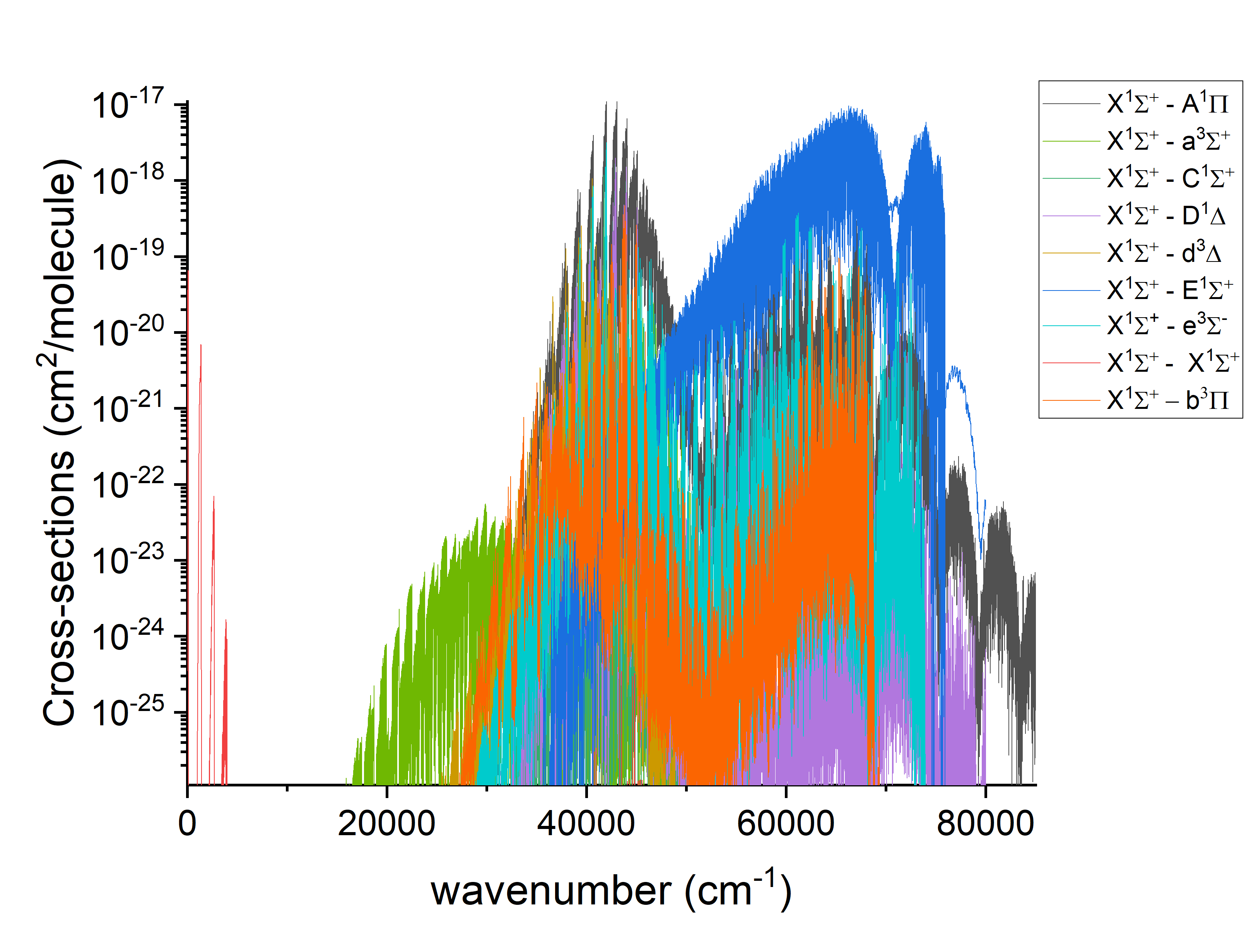}
    \caption{Overview of the calculated absorption spectra of PN from our model at $T = $ 2000~K. A Gaussian line profile of HWHM=1~\cm\ is used.}
    \label{f:T2000k}
\end{figure}

An overview of the PN absorption spectra in the form of cross sections at the temperature $T=$ 2000~K is illustrated in Fig.~\ref{f:T2000k}. Here, a Gaussian line profile with a half-width-at-half-maximum (HWHM)
of 1~\cm\ was used.  This figure shows contributions from each electronic band originating from the ground electronic state. The only band systems that have so far been characterized experimentally  are \X--\X\, \A--\X , \E--\X. Here we  mainly concentrate on discussing these band systems, due to the lack of experimental detection for other bands calculated in this work.

\subsubsection{\X\ band}

\begin{figure}
    \centering
    \includegraphics[width=0.7\columnwidth]{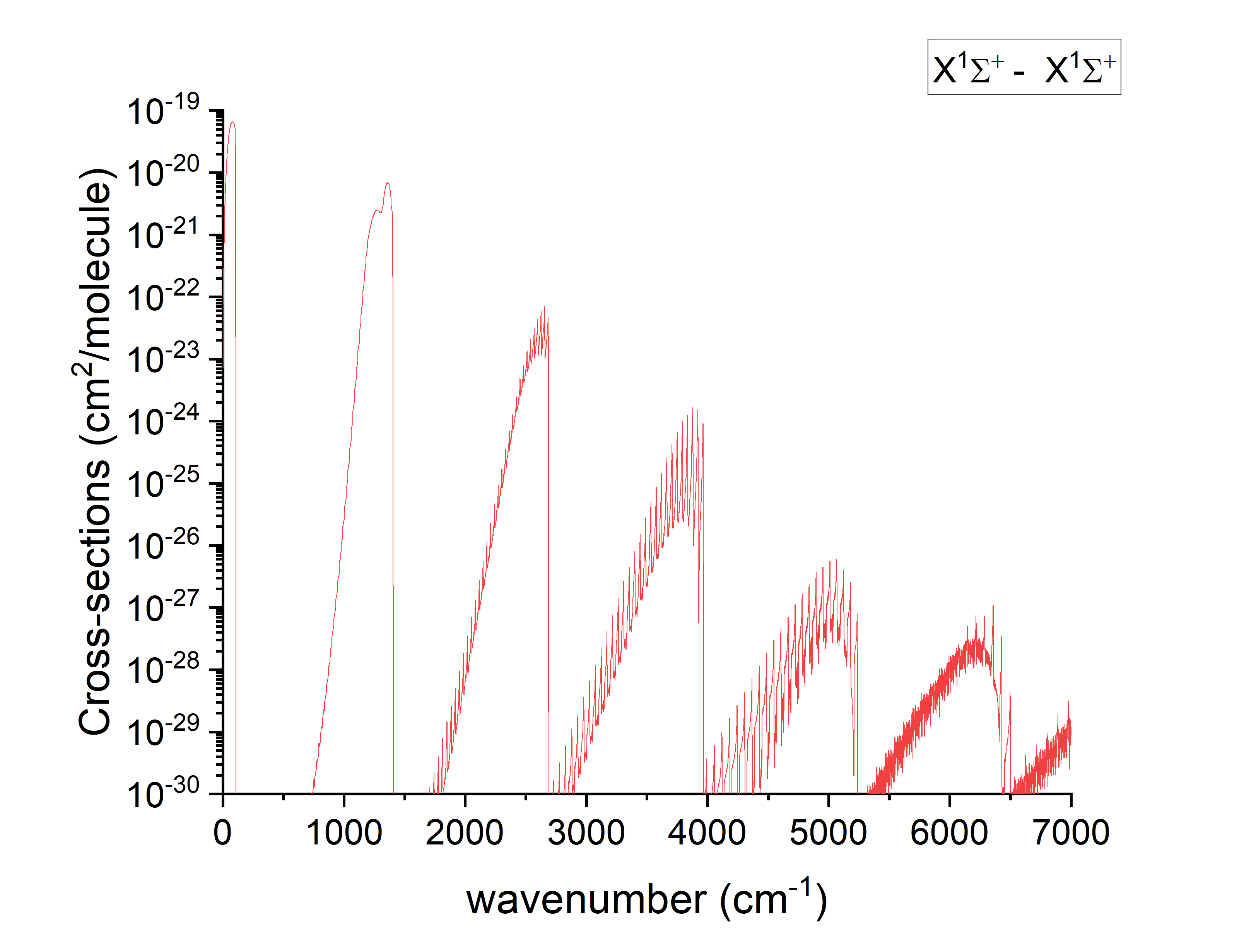}
    \caption{Overview of the calculated \X\ state rotation-vibration absorption spectrum of PN at $T= $ 2000~K with a Gaussian line profile of HWHM=1~\cm.}
    \label{f:X_2000K}
\end{figure}

The pure rotational $X$ state  transitions represent the main source of the PN observations  in interstellar-medium and stellar spectra.\citep{87TuBaxx.PN,87Ziurys.PN,90TuTsBa.PN,11YaTaSa.PN,16FoRiCa.PN,08MiHaTe.PN,13DeKaPa.PN,18ZiScBe.PN}
Apart from astrophysical observations, the band has been analysed in multiple lab experiments as well.\cite{77AtTixx.PN,72WyGoMa.PN,72HoTiTo.PN, 95AhHaxx.PN} An overview of the calculated absorption band of PN at 2000~K is shown in Figure~\ref{f:X_2000K}.

\subsubsection{\A--\X\ band}

\begin{figure}
    \centering
    \includegraphics[width=0.7\columnwidth]{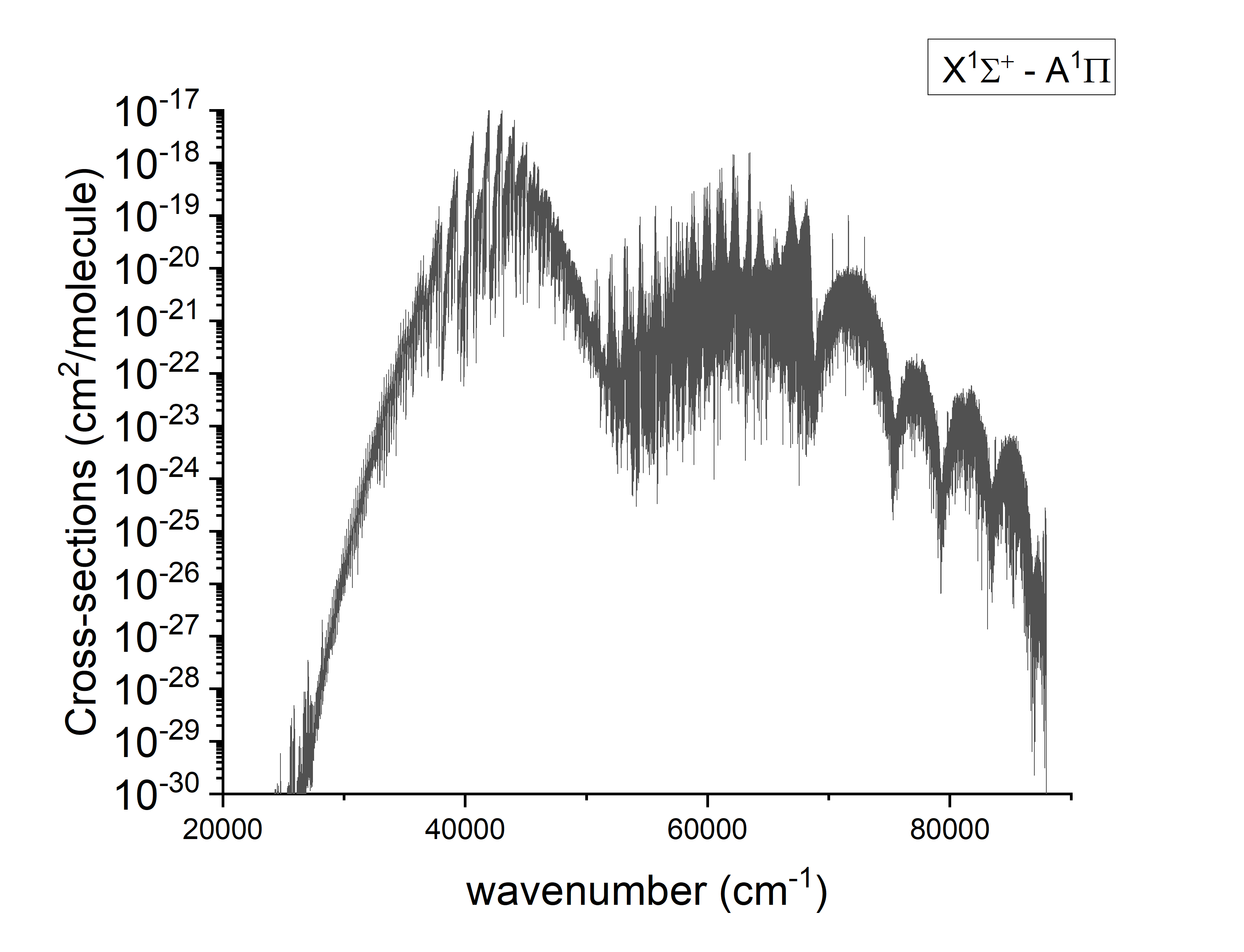}
    \caption{Calculated absorption spectra of the \A--\X system at 2000~K with a Gaussian profile of HWHM=1~\cm.}
    \label{f:A_2000K}
\end{figure}

The visible $A$--$X$ band  system was first observed by \citet{33CuHeHe.PN} and there have been  several subsequent laboratory observations.\cite{73MoSixx.PN,81GhVeVa.PN,87HeMaSt.PN,96LeMeDu.PN} Below we compare our results to  the experimental spectra; an overview spectrum for the \A--\X\ band system at 2000~K is given in Fig.~\ref{f:A_2000K}.

\begin{figure}
    \centering
    \centering
    \includegraphics[width=0.9\columnwidth]{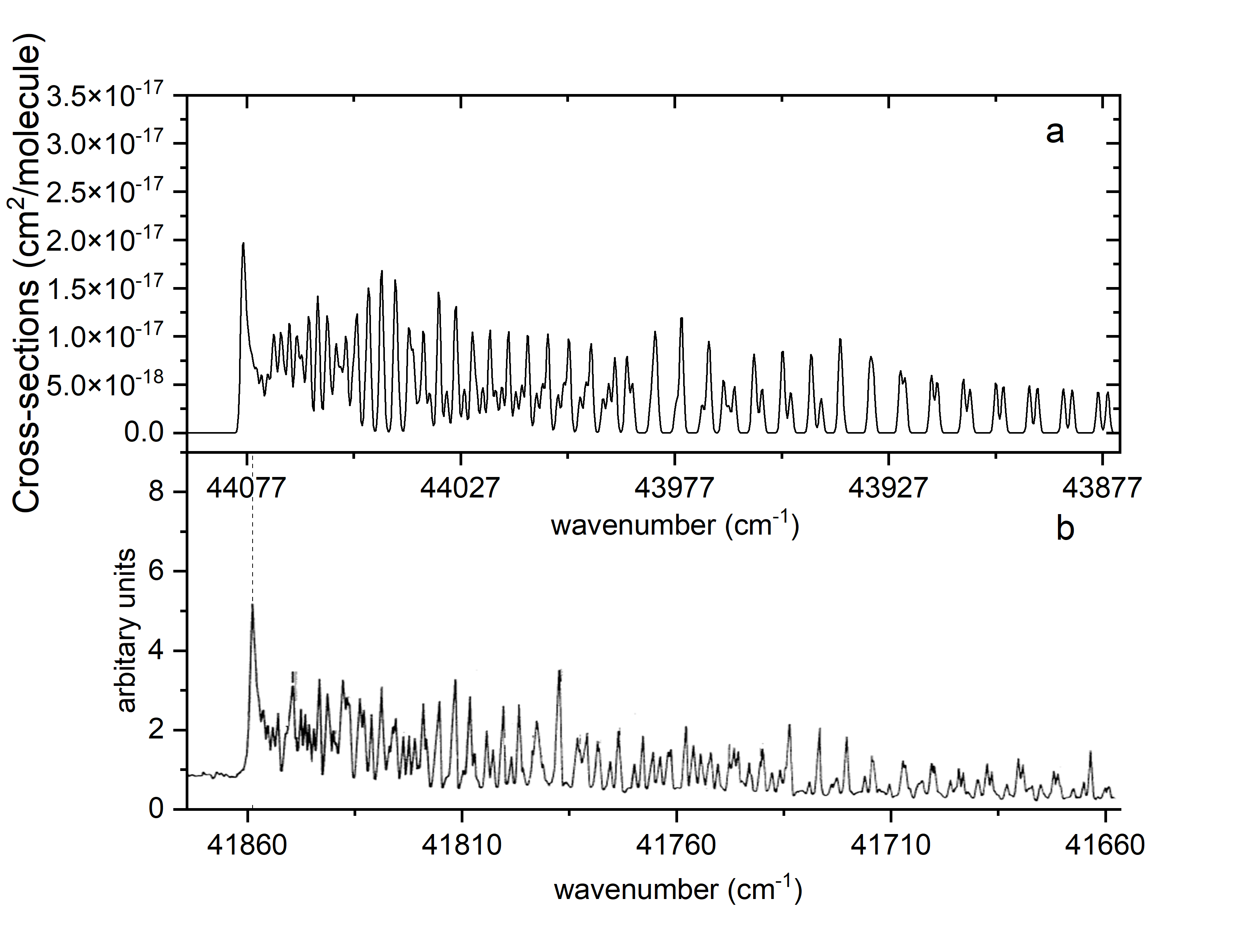}
    \caption{Comparison of the \A\ -- \X\ (2,0) simulated absorption band (panel a) with that recorded by \citet{96LeMeDu.PN} (panel b). The spectrum was simulated at $T= $1173.15~K  with a Gaussian profile of HWHM = 0.5 \cm\ and is offset by 2217~\cm\ so that the features align.}
    \label{f:96LeMeDu}
\end{figure}

Figure  \ref{f:96LeMeDu} compares a \Duo\ generated spectrum of the \A--\X\ (2,0) band to a part of the spectrum from recent experimental observations by \citet{96LeMeDu.PN}. Even at such a high resolution the main trends of the spectrum are reproduced. However, we have to note  a shift of  $-2217$ \cm\  and some differences in  intensities. This will be corrected with a further improvement of our model by fitting the \A\ PEC to the experimental data.

\begin{figure}
    \centering
    \includegraphics[width=0.99\columnwidth]{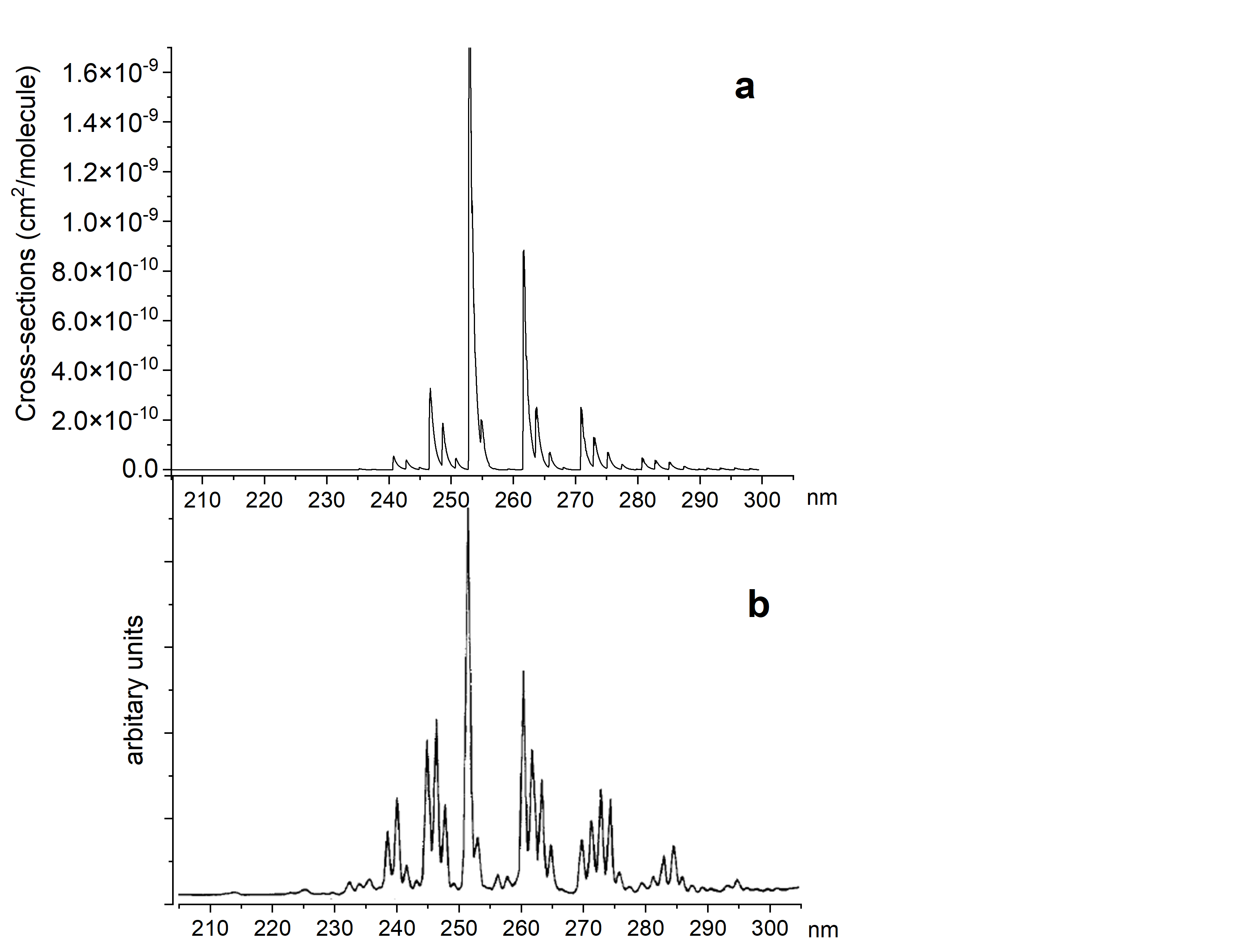}
    \caption{Comparison of \A -- \X simulated emission spectra (panel a) with that recorded by \citet{87HeMaSt.PN} (panel b). The spectra was simulated at 2000 K rotational temperature with a Gaussian profile and HWHM = 100 \cm}, 
    \label{f:87HeMaSt}
\end{figure}

Figure \ref{f:87HeMaSt} illustrates a simulated  emission  \A\ -- \X\ band system at 250~nm, which is compared to the chemiluminescent experimental spectrum of \citet{87HeMaSt.PN}, where we used the rotational temperature setting as 2000~K. However, when trying to adjust this spectrum, it proved to be very sensitive to the $T_e$ and $r_e$ values for the \A\ state. Therefore, for the purpose of this illustration  we shifted the \A\ PEC to match experimental values of $T_e$ and $r_e$.    
There is also a difference in relative magnitudes, which is attributed to the difference in vibrational population distribution of the experiment and our LTE calculations, as there are non-LTE effects which contribute to the spectrum.  Even though the PEC was adjusted to the experimental values previously reported by \citet{81GhVeVa.PN} to reproduce this spectrum, there is still a difference of 0.6~nm between \ai\ spectrum  and experiment, 
which we aim to resolve with further improvements of our spectroscopic model via empirical refinements.

\begin{figure}
    \centering
    \includegraphics[width=0.98\columnwidth]{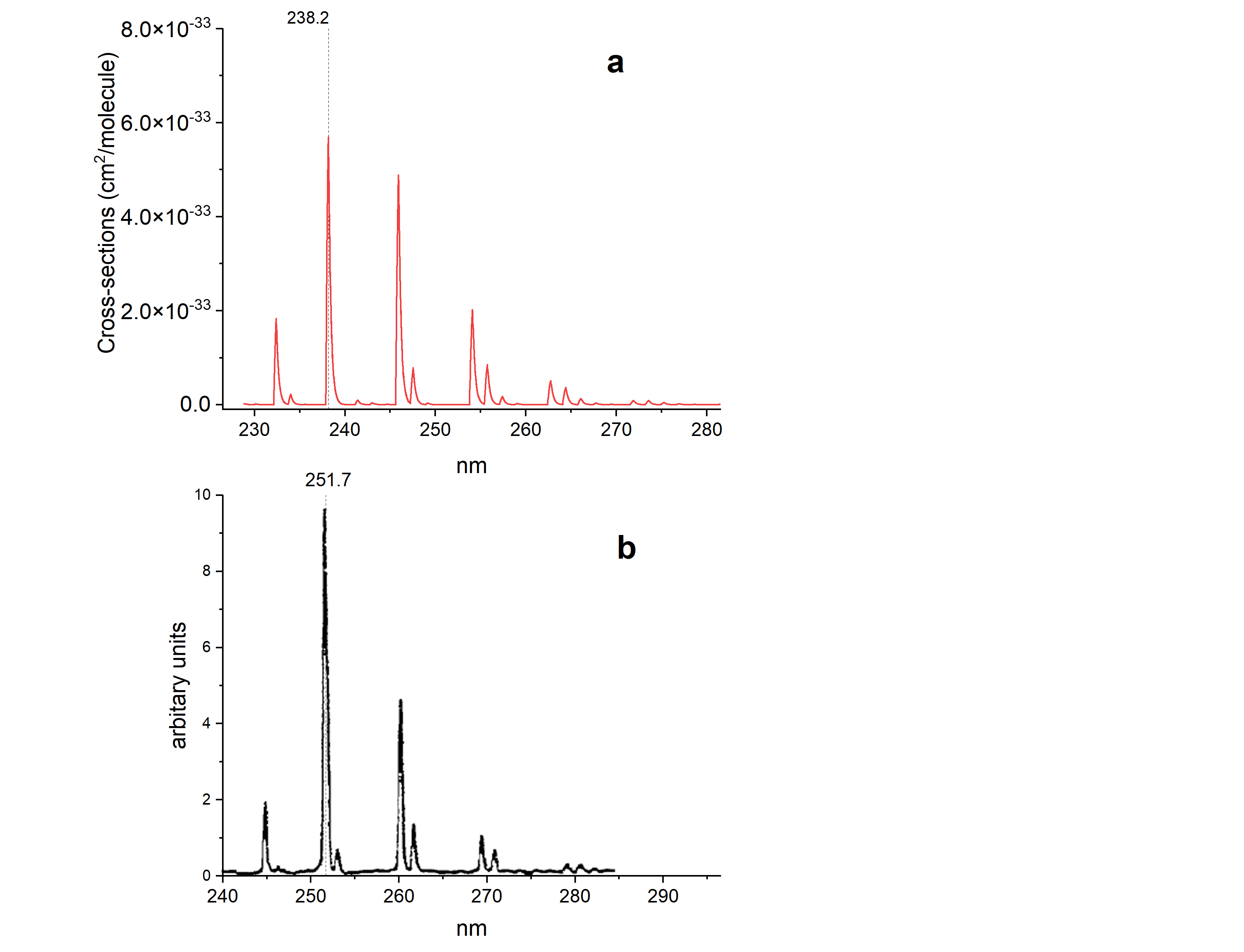}
    \caption{Comparison of an \A -- \X simulated emission spectrum (panel a) with that recorded by \citet{73MoSixx.PN} (panel b). The number at the top indicates the centre of the peak. The spectrum was simulated at  $T_{\rm rot}=$ 500~K (rotational) and $T_{\rm vib}=$ 1200~K (vibrational) temperatures with a Gaussian profile of HWHM = 100 \cm.}
    \label{f:73MoSixx}
\end{figure}

Figure~\ref{f:73MoSixx} shows a simulated non-LTE absorption  \A--\X\  spectrum of PN to compare  to the experiment of \citet{73MoSixx.PN}. Most of the major features are again shifted by $\sim$13.5~nm. We  attribute the relative difference in strength of peaks between theory and experiment to difficulty in reproducing the non-LTE behaviour of the experiment, due to the lack of necessary information in the original paper.

\subsection{Lifetimes}

\begin{figure}
    \centering
    \includegraphics[width=0.99\columnwidth]{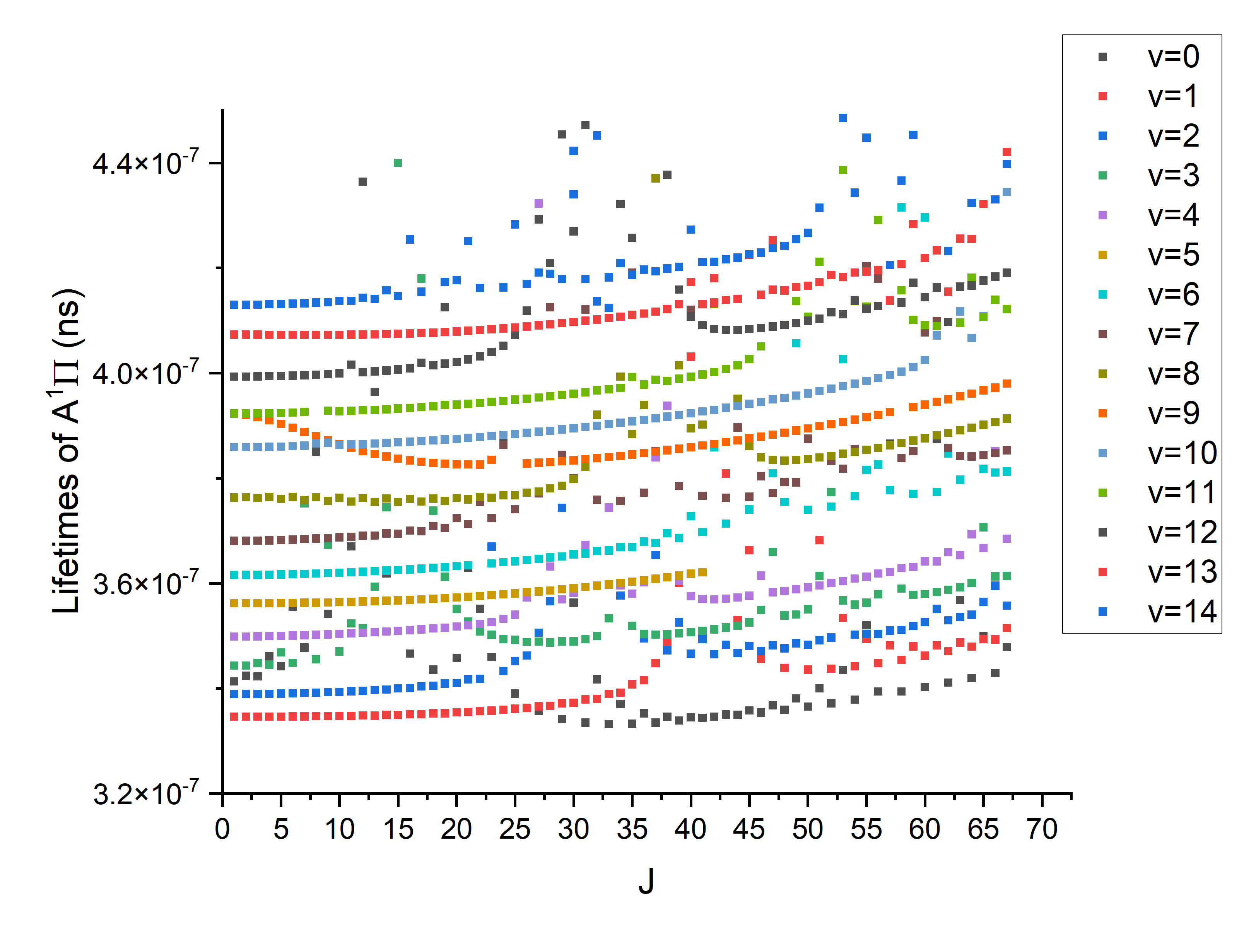}
   \caption{Lifetimes of \A\ state vs J. The top row of points is $v=14$ whereas the bottom is $v=0$}
    \label{f:lifetimes}
\end{figure}

In this work, lifetimes of PN are calculated using {\sc ExoCross}\cite{jt708}, based on the  states and transitions files generated by \Duo. ExoCross calculates lifetimes as follows:
\begin{equation}
      \tau_{i}=\frac{1}{\sum_{j<i}{A_{ij}}}
\end{equation}
where $\tau_i$ is radiative lifetime, $A_{ij}$ is Einsteins A coefficients, and $i$ and $j$ stand for upper and lower states, respectively. While the lifetimes reported in other works cited below are for vibrational levels and the above formula is for an individual state; however we have found no strong $J$-dependency for the PN lifetimes. This means that $J=0$ does indeed give a good approximation for the vibrational state lifetime. The lack of this dependency can be seen in  Fig.~ \ref{f:lifetimes}. The methodology used is described in detail by \citet{jt624}. Figure~\ref{f:lifetimes} shows  structures for different vibrational levels, which speaks to the perturbed nature of the \A\ state.

The lifetimes of $^{31}$P$^{14}$N in the \A\ state for ($v = 0$) were measured by \citet{75MoMcSi.PN}, and also were calculated by \citet{19QiZhLi.PN} and \citet{93deFeKo.PN}. While \citet{75MoMcSi.PN} report a lifetime of 227 $\pm$ 70 ns for $v'=0$ of \A\, our value is only slightly higher at 341 ~ns, which is closer than previous theoretical calculations of 695.4~ns\cite{19QiZhLi.PN} and 742.4~ns\cite{93deFeKo.PN} by a factor of 2. Table~\ref{t:lifetimes} provides a more detailed comparison between our calculated lifetimes and those of \citet{19QiZhLi.PN}. We assume that the major differences in the \A\ lifetimes and previous calculation to the difference in the transition dipole moment between \X\ and \A\, comparison of which can be seen in Fig.~\ref{f:A1Pi_TDMC_comparison}.
In order to check that, we have extracted the \A\ TDMC data from the original source \citep{19QiZhLi.PN} and rerun the lifetime calculation, then getting a comparable result of 677~ns for $v'=0$.
For the \bp\ lifetimes seem to be in agreement with \citet{19QiZhLi.PN} for all but v=0, but same cannot be said for the \D\ and \es\ states. We attribute the bulk of differences in lifetimes for these states to the differences in TDMCs.

Lifetimes for other singlet and triplet states are reported in Table~\ref{t:other_lifetimes}. From there we can see that the radiative lifetimes are in milliseconds for \C\ \dd\ and \as, and nanoseconds for \E. Long lifetimes for the \C, \dd\ and \as\ states are an indication of low probability of transition from these states, leading us to believe that these states would be very difficult to observe in an experiment. This is confirmed with current experimental evidence for PN, as \C, \D, \as, \dd, \es, which have only been observed indirectly through perturbation with the state \A\ by \citet{96LeMeDu.PN}. The \bp, \dd, \es, states have been also previously identified only through perturbation with \A\ state by \citet{87SaKrxx.PN}.

\begin{table*}
\caption{Comparison of lifetimes from our current work (A) and \citet{19QiZhLi.PN} (B). An experimental lifetime 227 $\pm$ 70 ns  was reported by  \citet{75MoMcSi.PN} for $v'=0$ of \A.}
\begin{tabular}{rllllllll}
    \hline
    \hline
   & \multicolumn{2}{l}{\A /ns}   & \multicolumn{2}{l}{\D / $\mu$s }   & \multicolumn{2}{l}{\bp / $\mu$s}  & \multicolumn{2}{l}{\es / $\mu$s}  \\
      \hline
$v'$ & A & B & A & B & A & B & A & B  \\
    \hline
0  & 341.3 & 659.4& 936.03& 3872 & 18.76 & 49.2 & 128.51 & 206.9\\
1  & 334.6 & 674.3& 301.51& 1186 & 15.38 & 27.81& 52.43 & 98.46\\
2  & 338.9 & 660.8& 120.96& 615.2& 13.07 & 19.5 & 73.08 & 59.45\\
3  & 344.3 & 646.5& 110.39& 378.3& 11.43 & 15.05& 61.85 & 40.75\\
4  & 349.8 & 642.2& 88.39& 246.8& 10.22 & 12.4 & 53.48 & 30.41\\
5  & 356.2 & 643  & 69.23& 181.8& 10.06 & 10.62& 47.30 & 23.68\\
6  & 361.6 & 632.1& 56.85 & 144.3& 8.55 & 9.364& 42.60 & 19.26\\
7  & 368.1 & 610.6& 47.37& 113 & 7.94 & 8.356& 38.88 & 16.16\\
8  & 376.3 & 607.4& 40.67 & 9.37 & 7.45 & 7.458&35.84 & 13.93\\
9  & 392.4 & 598.6& 35.52 & 9.85 & 7.04 & 6.881& 33.40 & 12.2 \\
10 & 399.3 & 583.8& 31.58 & 10.31 & 6.70 & 6.232& 31.33 & 10.77\\
11 & 407.3 & 581.8& 28.33 & 11.32 & 6.42 & 5.592& 29.60 & 9.517\\
12 & 413.0 & 566.8& 25.69 & 13.06 & 6.19 & 5.283& 28.07 & 8.146\\
13 & 420.7 & 565.3& 23.34 & 17.18 & 5.99 & 4.819& 26.78 & 7.142\\
14 & 431.3 & 553.1& 21.66 & 22.57 & 5.82 & 4.303& 25.67 & 6.192\\
    \hline
    \hline
\end{tabular}

\label{t:lifetimes}
\end{table*}

\begin{table}
    \centering
    \caption{Calculated lifetimes for the \C\ , \as\  and \dd\, \E\ states with  $v'$ up to 14. A full list of lifetimes is included in the Exomol states file.}
    \begin{tabular}{ccccc}
    \hline
    \hline
        $v'$ &\C /ms & \as /s &\dd /ms &\E /ns\\
        \hline
        0 & 65.02& 2.07$\times10^{6}$& 4.19 & 58.70\\
        1 & 53.72& 9.58& 9.81& 52.43\\
        2 & 19.31& 4.54& 5.86& 48.71\\
        3 & 8.21& 2.85& 2.77& 44.25\\
        4 & 4.35& 2.02& 1.44& 39.88\\
        5 & 2.68& 1.50& 0.64& 37.44\\
        6 & 1.83& 1.18&  0.03& 34.03\\
        7 & 1.33& 0.97&  0.39& 30.09\\
        8 & 1.02& 0.81&  0.19&28.67\\
        9 & 0.81& 0.69&  0.35& 27.89\\
        10 & 0.66& 0.06& 0.31& 27.76\\
        11 & 0.56& 0.41&  0.27& 27.97\\
        12 & 0.48& 0.33&  0.23& 28.23\\
        13 & 0.42& 0.09&  0.17& 28.71\\
        14 & 0.37& 0.04& 0.05& 29.77\\
    \hline
    \hline
    \end{tabular}
     \label{t:other_lifetimes}
\end{table}

\section{Conclusion}

In this work, a comprehensive \ai\ spectroscopic model for the nine lowest electronic states of PN is presented. A full set of potential energy, (transition) dipole moment, spin-orbit coupling, and electronic angular momenta coupling curves for these  9 electronic states  was produced \ai\ using the  icMRCI+Q/ECP10MWB and icMRCI/aug-cc-pV5Z(-DK) methods. These curves were then processed via the \Duo\ program  to solve the fully-coupled nuclear-motion Schr\"{o}dinger equation. Many of the results show satisfactory agreement with previous computational works, but there are certain differences in lifetimes and predicted spectra which indicate that further investigation backed by experimental data is needed to obtain a reliable spectroscopic model. In the next work we aim to tackle such differences, by producing an accurate, empirical line list for $^{31}$P$^{14}$N for use in astrophysical spectroscopy of distant stars and exoplanets. In order to achieve this, the \ai\ curves will be refined by fitting to the experimental data collected and processed via MARVEL methodology \cite{MARVEL}. Additionally non-adiabatic coupling effects, which are most important for the heavily perturbed \A\ state, will be included for finer accuracy. With  PN being  detected in multiple different interstellar media and phosphorous being a key element to life as we know it, PN is becoming a more important spectroscopic molecule. This work should provide an improved data quality for spectroscopic searches of PN in different astronomical environments, potentially leading to the detection of PN at ultraviolet
wavelengths.

The supplementary materials in this paper contain our  final spectroscopic model for PN in a form of a \Duo\ input file, an ExoMol States file including lifetimes and a partition function file.  This \Duo\ file is ready to be directly used with the \Duo\ program, see \url{http://exomol.com/software/}  allowing our results to be reproduced directly.

\section*{Acknowledgements}

The authors thank Alec Owens for helpful discussions during this project.  This work was supported by UK STFC under grant ST/R000476/1. This work made  use of the STFC DiRAC HPC facility supported by BIS National E-infrastructure capital grant ST/J005673/1 and STFC grants ST/H008586/1 and ST/K00333X/1. We thank the European Research Council (ERC) under the European Union’s Horizon 2020  research and innovation programme through Advance Grant number  883830. We also want to thank Moscow Witte University for sponsoring the fellowship enabling this research. This work is additionally supported by  Khalifa University of Science and Technology under Award No. CIRA-2019-054. Khalifa University High power computer was used for the completion of this work.



\providecommand*{\mcitethebibliography}{\thebibliography}
\csname @ifundefined\endcsname{endmcitethebibliography}
{\let\endmcitethebibliography\endthebibliography}{}

\end{document}